\documentclass[a4paper,11pt]{article}

\usepackage{jheppub} 

\usepackage[T1]{fontenc} 

\title{Entanglement entropy and complexity of singular subregions in deformed CFT }

\author[a]{E. Bakhshaei,}
\author[a,b]{A. Shirzad,}

\affiliation[a]{Department of Physics, Isfahan University of Technology,\\P.O.Box 84156-83111, Isfahan, Iran}
\affiliation[b]{School of Particles and Accelerators,\\Institute for Research in Fundamental Sciences
(IPM), P.O.Box 19395-5531, Tehran, Iran}

\emailAdd{e\_bakhshae@yahoo.com}
\emailAdd{a.shirzad.iut@gmail.com}

\abstract{In the framework of the AdS/CFT correspondence, imposing a scalar field in the bulk space-time leads to deform the corresponding CFT in the boundary, which may produce corrections to entanglement entropy, as well as the so-called subregion complexity. We have computed such corrections for a set of singular subregions including kink, cones and creases in different dimensions. Our calculations shows new singular terms including universal logarithmic corrections for entanglement entropy and subregion complexity for some distinct values of conformal weight.}

\keywords{AdS/CFT correspondence, Entanglement entropy, Subregion complexity}

\arxivnumber{1902.02504 }

\begin{document} 
\maketitle
\flushbottom

\section{Introduction}
\label{intro}
Entanglement entropy of a subregion in a conformal field theory is conjectured to be found holographically as the surface of the extension of subregion into the bulk (i.e. RT surface) provided that its variation vanishes \cite{Ryu:2006bv,Ryu:2006ef}. This quantity diverges near the boundary. Most of the time people are interested in the behavior of the divergent terms and their relationship with physical properties of the system. Especially the logarithmic singularity may be interpreted as some cut-off independent characteristics of the system. For instance, in three dimensions the coefficient of  logarithmic term has been shown to be proportional to the central charge of the  corresponding CFT at the boundary \cite{Casini.2007,Casini.2009,Takayanagi.2007, Moore.2006}. Such relationship are also aimed for higher dimensions in subsequent works \cite{Solodukhin:2008cdf}.

There is also some interest in the entanglement entropy of singular subregions  in recent years \cite{Bueno:2015rda,Bueno:2015xda,mollabash12,mollabash13}. The effect of singularities gained some attractions, first in three dimensions \cite{Takayanagi.2007} and then in higher dimensions \cite{Singh:2012ala}. The most interest is to find new singular terms in the expansion of the entanglement entropy due to singularity of the considered subregion \cite{Myers:2012vs}.

Another quantity of great interest is the complexity of a QFT living in the boundary. Besides the well-known approaches of $ complexity = volume $ \cite{Stanford:2014jda, Susskind:2014rva,Susskind:2014moa,Couch:2016exn} and $ comlexity =action $ \cite{Carmi:2016wjl, Lehner:2016vdi,Brown:2015bva,Brown:2015lvg}, there is also another conjecture to  consider the complexity as the volume enclosed by the Rio-Takayanagi surface \cite{Alishahiha:2015rta,Ben-Ami:2016qex}.This approach is known as subregion complexity. 
In a recent paper \cite{Bakhshae:2017} we used this approach to study the subregion complexity of a number of singular surfaces, focusing on the singularities in terms of UV cutoff parameter. 

Recently some interests have been arised to deform the CFT by imposing a relevant operator \cite{parvizi1,parvizi3,parvizi4,parvizi5,parvizi6,parvizi7}.
Using the standard AdS/CFT correspondence \cite{myers11}, this may be achieved, for example, by turning on a scalar field in the bulk. Hence, the geometry of the bulk is no longer pure AdS; however, it turns out to be asymptotically AdS near the boundary. It has been shown that this deformation may lead to appearing universal logarithmic correction in the entanglement entropy for definite values of the conformal dimension of the relevant operator \cite{myers}.
The effect of deformation of the CFT on the entanglement entropy is 
investigated for regular subregions such as sphere in Refs. \cite{parvizi6,myers}. For singular subregions, this effect has  been recently studied only for kink in three dimensions in \cite{prvizi}. Some general aspects of the complexity of the deformed theories are studied in \cite{alishahiha11}.  

In this paper we are mostly interested in calculating the entanglement entropy and complexity for a set of singular surfaces (including cones and creases) in a deformed conformal field theory.
After a brief review of a deformed conformal field theory in the next section, we calculate the entanglement entropy for singular subregions of a kink in $d=3$ and cones in $d=4, 5$ and $6$ for a deformed theory. We also consider creases in  $d=4, \ 5$ and $6$. This is done in section~\ref{Entanglement Entropies}. The essential calculations and technical points are given in more details for the case of kink and to some extent for cone $c_1$. For other singular surfaces we give only the important results.
In section~\ref{Complexity} we give our results for the subregion complexities of the same singular submanifolds. We discuss about our results in section~\ref{Discussions}.

It is important to distinguish among different types of singularities. The most familiar kind is that of ordinary UV divergences of the entanglement entropy, as well as the complexity, when we approach the boundary in the AdS/CFT framework. The next kind corresponds to geometrical singularities near the needle points or wedges of singular subregion. Finally we encounter new singular terms due to deforming a theory by a relevant operator.  Among different singular terms people are mostly interested in the logarithmic singular terms because of their universal characteristics due to independence of the regularization process. 
\label{Introduction}

\section{Deformed CFT}
\label{Deformed CFT}
Consider a CFT living in the $d$-dimensional boundary of a $(d+1)$-dimensional space-time. As is well-known \cite{myers}, turning on a scalar field in the bulk, one can deform  the CFT by a relevant operator. To do this, the Hilbert-Einstein action with a negative cosmological constant $\Lambda=-\dfrac{d(d-1)}{L^2}$ is perturbed as follows
\begin{align}
I=\dfrac{1}{16\pi G_N}\int d^{d+1}x\sqrt{g}\left[R-\dfrac{d(d-1)}{L^2}\right]-\frac{1}{2}\int d^{d+1}x\sqrt{g}\left[(\partial \Phi)^2+M^2 \Phi ^2\right].
\label{a50}
\end{align}
The mass parameter $ M$ determines the conformal dimention of the boundary operator $O$ (dual to $\Phi$ ) as
\begin{align}
\Delta _\pm=\frac{d}{2}\pm \sqrt{(ML)^2+\frac{d^2}{4}},
\label{a51}
\end{align}
where $\Delta _+$ and $\Delta_-$ are valid for $ -\frac{d^2}{4}\leq (ML)^2 \leq-\frac{d^2}{4}+1 $;  while  for $ -\frac{d^2}{4}+1< (ML)^2 $ only $\Delta _+$ is valid \cite{nishi1}. Due to this deformation the AdS solution of the Einstein equation also deforms as follows
\begin{align}
ds^2=\dfrac{L^2}{z^2}\left[\dfrac{dz^2}{f(z)}-dt^2+d\rho ^2+ \rho^ 2(d\theta ^2+ \sin ^2\theta d\Omega _n ^2)\right], 
\label{a52}
\end{align}
where $f(z)$ is the deformation function. The geometry is asymptotically AdS, i.e. $ f(z)\rightarrow1$ as $z\rightarrow0$. Considering the equations of motion for metric components as well as the scalar field $\Phi$, one can show directly that the expansion of $ f(z)$ near the boundary is as follows
\begin{equation}
f(z)=\left\{ \begin{array}{ll}
	1+ (\mu z)^{2\alpha}+..., \quad &\Delta \neq d/2\nonumber\\
( \mu z)^{d}(\log \mu z)^2+..., \quad &\Delta = d/2 \end{array} \right.
\end{equation}
where $\mu $ is a mass parameter determined from the parameter $\lambda$ of coupling  the relevant operator, and 
$\alpha = d-\Delta_+=\Delta_-$ for  both $\Delta=\Delta_+$ and $ \Delta=\Delta_-$.

Deforming the bulk geometry leads to some changes in the entanglement entropy of the subregions as stated above. It is shown \cite{parvizi6} that to lowest order of $\mu$, the variation of the entanglement entropy for the sphere in a $d$-dimensional CFT  and $ \Delta \neq d/2$ reads
 \begin{align}
 \delta S=-\dfrac{\Gamma(\dfrac{d+1}{2})\Gamma(\dfrac{2-d+2 \alpha}{2})}{4\Gamma(\frac{3}{2}+\alpha)}K(\mu R)^{2\alpha}+\dfrac{K(\mu R)^{d-2}}{2(2-d+2 \alpha)}(\mu\delta)^{2-d+2\alpha}+\mathcal{O}(\delta ^{4-d+2\alpha}),
  \end{align}
 where $K=L^{d-1} \textstyle{Vol} (S^{d-2}) /4 G_N $, $\delta$ is the UV cutoff and $R$ is the radius of the sphere. The first term is the finite result while the subsequent terms are singular corrections. They are a limited number of terms which are present unless $2n-d+2\alpha \le 0$ where equality leads to logarithmic singularity. For $\Delta=\Delta_-$ where $d/2-1<\alpha<d/2$, we find $n<1$ which gives no contribution for $\alpha \ne d/2$.  For $\Delta=\Delta_+$ we have $d/2<\Delta_+<d$ which leads to $0<\alpha<d/2$. Hence, for $d=3$ we have a singular term as $\delta^{-1+2\alpha}$ for $\alpha<1/2$ a logarithmic singularity for $\alpha=1/2$ and no singular term for $\alpha>1/2$. For $d=4, 5$ and $6$  the results are consistent with what we will find in the next section (see the table~\ref{tab1} ).
 \section{Entanglement Entropies}
\label{Entanglement Entropies}
Let us rewrite the solution \eqref{a52} of the equations of motion of the action \eqref{a50} in the following form
\begin{align}
	ds^2=\frac{L^2}{z^2}\left[ - dt^2+\frac{dz^2}{f(z)}+d\rho ^2+\rho ^2 (d\theta ^2+ \sin^2 \theta d \Omega _n ^2) +\sum _{i=1}^{i=l}d x_i ^2\right],\label{a0}
\end{align}
In this form we have divided the flat $(d-1)$-dimensional manifold of the boundary at fixed time into a $(n+2)$ flat space described by spherical coordinates and a $l$-dimensional space described by Cartesian coordinates, such that $d=l+n+3$. This enables us to introduce the geometrical singularity by assuming $ \rho\rightarrow 0 $  when the angle $\theta$ is limited to the interval $[-\Omega,\Omega]$ for kink (i.e. $n=0$ and $l=0$) and the interval $[0,\Omega]$ for cones (i.e. $n \ge 1$ and $l=0$). Creases correspond to extensions $l \ge 1$. All of our singular subregions are restricted to $\rho \le H$ where $H$ is the IR cutoff.
Assuming $\rho=\rho(z, \theta)$ to describe the RT surface, the induced metric reads 
\begin{equation}
h=\left(\begin{array}{c c c c c c}
\dfrac{L^2}{z^2}(\frac{1}{f(z)}+\rho '^2)&\dfrac{L^2}{z^2}\rho' \dot{\rho} &&&&\\
\dfrac{L^2}{z^2}\rho' \dot{\rho}&\dfrac{L^2}{z^2}(\rho ^2+\dot{\rho}^2)&&&&\\
&&\dfrac{L^2\rho ^2 \sin ^2(\theta)}{z^2 }g_{ab} (S^n)&&&\\
&&&\dfrac{L^2}{z^2}&&\\
&&&&\ddots &\\
&&&&&\dfrac{L^2}{z^2} \\ 
\end{array}\right).
\label{ab11}
\end{equation}
According to RT prescription, the entanglement entropy is proportional to the minimized  area of the RT surface as 
\begin{equation}
S=\frac{2\pi}{l_p^{d-1}} L^{d-1} \Omega_n  \int dz  d\theta\,\dfrac{\rho^{n}}{z^{d-1}}\,\sin^n \theta\,\sqrt{{\rho ' }^{2}\rho ^2+\frac{\rho ^2 +\dot{\rho}^2}{f(z)}}, \label{ab10}
\end{equation}
where $ \Omega_n$ is the volume of the unit $n$-sphere,  $\dot{z}=\partial_ {\theta} z $ and $ z'=\partial _{\rho} z $. In the following subsections we  calculate the  entanglement entropy for different singular subregions.
\subsection{kink $ k $}
\label{kink }
\begin{figure}[tbp]
\centering 
\includegraphics[width=.45\textwidth]{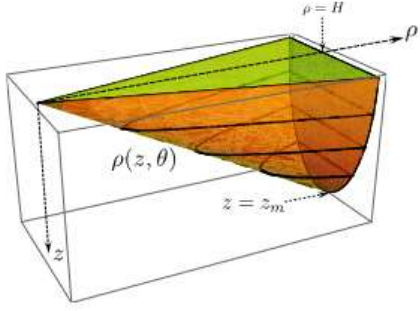}
\caption{\label{fig1}Schematic figure of the RT extension of the kink (green region) into the bulk \cite{Myers:2012vs}. }
\end{figure}

 The entanglement entropy \eqref{ab10} for kink is given by the following integral
\begin{align}
S\vert_{k}=\frac{2\pi L^2}{l_p ^2}\int d z d\theta L(\rho, \dot{\rho}, \rho'), \label{a'4}
\end{align}
where
\begin{align}
L(\rho, \dot{\rho}, \rho')= \dfrac{\sqrt{\frac{\dot{\rho}^2+\rho ^2}{f(z)}+{\rho '}^2 \rho^2 }}
{z^2}.\label{a4}
\end{align}
 Minimizing the above integral gives the equation of motion of $ \rho(z,\theta) $ as follows
\begin{align}
 2 f z \rho(\rho ^2 +\dot{\rho}^2){\rho} '' +2 z \rho(1+f{\rho ' }^2)\ddot{\rho}-4 f z \rho \rho ' \dot{\rho}\dot{\rho}'+z \rho \rho'(\rho ^2 +\dot{\rho} ^2) f '  & \nonumber\\
-2z\left((1+f{\rho '}^2)\rho ^2 +2\dot{\rho}^2\right)-4 f \rho \rho '\left((1+f {\rho '}^2)\rho ^2 +\dot{\rho}^2\right) & =0. \label{ab2}
\end{align}
This partial differential equation should be solved according to the following boundary conditions
\begin{align}
& \frac{\partial\rho}{\partial \theta }(z,0)=0,  \label{d2}\\ & \rho(z, \Omega) =H. \label{d3}
\end{align}
Eq. \eqref{d2} shows that at arbitrary $z$ the coordinate $\rho$ on RT surface acquires its minimum value at $\theta=0$, while Eq. \eqref{d3} shows that at the limiting points $\theta= \pm \Omega$ the RT surface touches the boundary $\rho=H$. See figure~\ref{fig1}
 to get a geometrical feeling about the problem.
To solve Eq. \eqref{ab2} we consider a perturbative ensatz where in the limit $ \mu \rightarrow 0 $, we assume $\rho=z/h(\theta)$. This leads to the following equation for $h(\theta)$
\begin{align}
2+3 h^2+h^4+2 \dot{h}^2+h(1+h^2)\ddot{h}=0,\label{a6}
\end{align}
which give the following constant 
\begin{align}
K_3=\dfrac{(1+h^2)}{h^2\sqrt{1+h^2+\dot{h}^2}}.\label{a10}
\end{align}

The boundary condition \eqref{d2} gives $\dot{h}(0)=0$ at the turning point $\theta=0$. Hence, the constant $K_3$ can be written in terms of $h_0=h(0)$. In principal the Eq. \eqref{a6} can be solved to find the function $h(\theta)$. However, we use this equation to find $\dot{h}$ as a function of $h$. Inserting these results in Eq. \eqref{ab2}, one can find the entanglement entropy of kink \cite{Myers:2012vs}. Then we can complete our enzatz by considering the following expansion 
\begin{align}
\rho (z,\theta)=\frac{z}{h(\theta)}+\mu^{2\alpha} z^{2\alpha+1} g_2(\theta)+\cdots .\label{a5}
\end{align}
Inserting the complete ansatz \eqref{a5} into the equation of motion\eqref{ab2} we find (in addition to Eq. \eqref{a6}) the following equations for $g_2$, 
\begin{align}
& 2 h^3(1+h^2)(1+h^2 +\dot{h}^2)\ddot{g}_2+2 h^2 \dot{h}(2(5 +2 h^2 +2 \alpha)\dot{h}^2+(1+h^2)(10+5 h^2+4 \alpha+h \ddot{h}))\dot{g}_2\nonumber\\
&-2h (-2(-2+h^2 +\alpha+2 \alpha ^2)\dot{h}^4+\dot{h}^2(22 +h^4 +2 (7 -2\alpha)\alpha+ h^2(19 +4\alpha- 8 \alpha ^2)\nonumber\\
&+2 h(2 + h^2 +2 \alpha)\ddot{h})+(1 + h^2)(2 (9 + 8 \alpha) + h^2 (19 + 4 h^2 - 4 (-2 + \alpha) \alpha)  \nonumber\\
&+3 h (2 + h^2 + 2 \alpha) \ddot{h})) g_2+(2 (-1 + \alpha) \dot{h}^4 + (1 + h^2) (-8 + h^4 + 
 2 h^2 (-2 + \alpha) \nonumber\\
&+ h (-2 + h^2)\ddot{h})+\dot{h}^2(h^4 + 2 (-5 + \alpha) + h^2 (-5 + 4 \alpha) + h (-1 + h^2)\ddot{h}))=0.\label{ab1}
\end{align}
As is expected, for the generic problem of the entanglement entropy, we will find UV divergences in the limit $z \rightarrow 0$ or equivalently as $\rho \rightarrow 0$ (for arbitrary $-\Omega <\theta <\Omega$) or $\theta \rightarrow \Omega$ (for finite $\rho$). In terms of the variable $h$ the latter limit is equivalent to $h \rightarrow 0$. The UV divergent terms of the entanglement entropy  originate from the divergences of the integrand as well as the limits of the integral  Eq. \eqref{a'4} as
\begin{align}
	S\vert_{k}=&-\frac{2\pi L^2}{l_p ^2}\int_{z_m} ^{\delta }dz \int_{-\Omega+\epsilon
	}^{\Omega -\epsilon} d\theta L(\rho, \dot{\rho}, \rho'), 
\end{align}
which upon changing the integration limits of $\theta$ from $(-\Omega,\Omega)$ to $(0, \Omega)$ and replacing $d\theta$ in Eq.  \eqref{a'4}
by $dh/\dot{h}$,  reads
\begin{align}
S\vert_{k}=&-\frac{4\pi L^2}{l_p ^2} \int_{z_m} ^{\delta }\frac{dz}{z} \int_ {h_0}^{h_{1c}} \frac{dh}{\dot{h}} L(\rho, \dot{h}d\rho/dh, \rho') . \label{aa8}
\end{align}
Concerning the integral bounds in Eq. \eqref{aa8},  from  the boundary condition \eqref{d3}  we have $\rho (z, \Omega -\epsilon(z)) = H $; hence we can define $ h_{1c}(z) = h(\Omega - \epsilon(z))$. In the limit $z=\delta \rightarrow 0$ we have  
$ h_{1c}(\delta) = h(\Omega - \epsilon(\delta))$. The limit $z_m$ is also achieved via the condition $\rho(z_m, 0) = H $. Inserting $\rho$ from Eq. \eqref{a5} into \eqref{aa8} gives the following expansion for the entanglement entropy with respect to $\mu$
\begin{align}
S\vert_{k}=-\frac{4\pi L^2}{l_p ^2}(I_0 +\mu ^{2\alpha} I_1+ \cdots ), \label{a9}
\end{align}
where
\begin{align}
&I_k=\int_{z_m} ^{\delta }dz z^{(2\alpha)k-1} \int_ {h_0}^{h_{1c}}dh G_k(h),\label{a11}
\end{align}
in which
\begin{align}
&G_0(h)=\dfrac{\sqrt{1+h^2 +\dot{h}^2}}{\dot{h} h^2},\label{aa10}\\
&G_1(h)=\dfrac{ -\dot{h}^2-h^2(1+2 \dot{h}\dot{g}_2)+4 (1+\alpha) h g_2+2 h^3 g_2}{2\dot{h} h^2 \sqrt{1+h^2+\dot{h}^2}}\, .\label{aa11}
\end{align}
Now to calculate the entanglement entropy one needs to solve Eq. \eqref{ab1}  for $ g_2 $. This is a difficult task. However, since we are mostly interested to find the UV divergent terms of the entanglement entropy, we just need to find the asymptotic behavior the corresponding function $g_2$ near the boundary, i.e. in the limit $h\rightarrow 0$. According to Eq. \eqref{a5} , in order to keep $\rho$ finite in the limit $ h \rightarrow 0 $ and $ \delta\rightarrow 0 $, the most singular term of $g_2$ should be of order $h^{-2\alpha-1}$. Assuming $g_2=a h^{-2\alpha-1}+a' h^{-2\alpha}+\cdots$ and inserting it in the corresponding equations \eqref{ab1} (as well as $\dot{g_2} =\dot{h}dg_2/dh$ etc.) and using Eq. \eqref{a10} for $\dot{h}$ we find the following results
\begin{align}
g_2=\dfrac{a_1}{h^{2\alpha +1}}+\dfrac{a_2}{h^{2\alpha-1}}+\dfrac{a_3}{h^{2\alpha-3}}+\frac{a_4}{ h}+a_5 h+\mathcal{O}(h^3),\label{a13}
\end{align}
where
\begin{align}
&a_2=\frac{a_1}{5} (2 + 5 \alpha - 2 \alpha^2),\\
&a_3=\frac{a_1}{70} (-12 + 10 K_3^2 - 7 \alpha+ 43 \alpha^2 - 28 \alpha ^3 + 4 \alpha^4),\\
&a_4=-\frac{1}{(6 + 4 \alpha)}\ ,\\
&a_5= \frac{2}{15 + 31 \alpha + 20 \alpha^2 + 4 \alpha^3}\ .
\end{align}
The constant $a_1$ would be fixed from the boundary conditions \eqref{d2} and \eqref{d3}.
Now we need to determine the limiting value $h_{1c}$ to find the divergencies of the integrals \eqref{a11}. This can be down from the same boundary condition $\rho(z, \Omega)=H$ in the limit $ z=\delta$, i.e. $\rho (\delta, \Omega-\epsilon)=H$ where $h_{1c} = h(\Omega-\epsilon)$. Using Eqs. \eqref{a10} and  \eqref{a13} we can find the expansion of  $ h_{1c} $ in terms of $ \delta $ as follows
\begin{align}
h_{1c}(\delta)=&\left( \frac{1}{H}+a_1 \mu ^{2\alpha} H^{2\alpha -1}\right)\delta +a_2\mu ^{2\alpha} H^{2\alpha -3}\delta ^3\nonumber\\ &+a_3 \mu ^{2\alpha} H^{2\alpha-5}\delta ^5+ \mu ^{2\alpha}\frac{a_4}{H}\delta ^{2\alpha+1} +\mu ^{2\alpha} \frac{a_5}{H^3}\delta ^{2\alpha+3}
+\mathcal{O}(\delta ^{2\alpha+5}).\label{a155}
\end{align}
From the Eqs. \eqref{a10} and \eqref{a13} we can find the following expansions for $G_0$ and $G_1$  in terms of the UV cut-off parameter $\delta$
{       
\begin{align}
&G_0(h)\sim- \frac{1}{h^2}-\dfrac{k_3 ^2 h^2}{2}+\mathcal{O}(h^4)\label{m1},\\
&G_1(h)\sim -\frac{a_1(1+2\alpha)}{h^{2\alpha+2}}+\frac{2+\alpha}{(3 +2 \alpha) h^2}+\mathcal{O}(h^{-2\alpha}).\label{m2}
\end{align}
Separating the singular part of the integrals $ I_k $, we can divide them as $I_k=I_k'+I_k'' $, where the integrand of $I'_k$ is regular. Hence, we have
\begin{align}
I_0=&I_0'+I_0''\nonumber\\
=&\int_{z_m} ^{\delta}\frac{dz}{z}  \int _{h_0}^{h_{1c}} dh\left(G_0(h)+\frac{1}{h^2}\right)-\int_{z_m} ^{\delta}\frac{dz}{z}  \int _{h_0}^{h_{1c}} dh\frac{1}{h^2}\nonumber\\
=&\int_{z_m} ^{\delta}\frac{dz}{z}  \int _{h_0}^{h_{1c}} dh\left(G_0(h)+\frac{1}{h^2}\right)+\int_{z_m} ^{\delta}\frac{dz}{z}\left(\frac{1}{h_{1c}}-\frac{1}{h_0}\right),
\label{aa15}
\end{align}
\begin{align}
I_1&=I_1'+I_1''\nonumber\\
=&\int_{z_m} ^{\delta}z^{2\alpha-1} dz\int _{h_0}^{h_{1c}} dh\left(G_1(h)+\frac{a_1(1+2\alpha)}{h^{2\alpha+2}}-\frac{2+\alpha}{(3 +2 \alpha) h^2}\right)\nonumber\\
&-\int_{z_m} ^{\delta} z^{2\alpha-1}dz\int _{h_0}^{h_{1c}}dh\left(\frac{a_1(1+2\alpha)}{h^{2\alpha+2}}-\frac{2+\alpha}{(3 +2 \alpha) h^2}\right) .
\label{a16}
\end{align}
 Note that for instance in the second term of Eq. \eqref{a16} due to the term $  z^2 dz $, we just need to consider the first two terms of the expansion of $ G_1 $ in Eq. \eqref{m2}. The same task is done in the other calculations. In order to find the singular behavior in each case, let take the derivative of $I_k$'s with respect to $ \delta $, i.e.
\begin{align}
\frac{d I'_0}{d\delta}=&\frac{1}{\delta } \int _{h_0}^{h_{1c}} dh\left(G_0(h)+\frac{1}{h^2}\right)\nonumber\\
&=\frac{1}{\delta } \int _{h_0}^{0} dh\left(G_0(h)+\frac{1}{h^2}\right)+\frac{d h_{1c}}{d\delta}\left(G_0(h)+\frac{1}{h^2}\right)\vert _{h=h{1c}}+..\nonumber\\
&=\frac{1}{\delta } \int _{h_0}^{0} dh\left(G_0(h)+\frac{1}{h^2}\right)+\frac{d h_{1c}}{d\delta}\left(-\frac{k_3 ^2 h_{1c}^2}{2}\right)\nonumber\\
&=\frac{1}{\delta } \int _{h_0}^{0} dh\left(G_0(h)+\frac{1}{h^2}\right)+\mathcal{O}(\delta ^2),\label{a17}
\end{align}
\begin{align}
&\frac{d I_0''}{d\delta}=\frac{H}{\delta ^2}\left(1-a_1 \mu ^{2\alpha} H^{2\alpha }-\mu ^{2\alpha} a_4\delta ^{2\alpha}\right)-\frac{1}{h_0 \delta }+\mathcal{O}(\delta ^0),\label{a18}\\
&\frac{d I_1'}{d\delta}=\delta ^{2\alpha-1} \int _{h_0}^{0} dh\left(G_1(h)+\frac{a_1(1+2\alpha)}{h^{2\alpha+2}}-\frac{2+\alpha}{(3 +2 \alpha) h^2}\right)+\mathcal{O}(\delta ^0 ),\label{a19}\\
&\frac{d I_1''}{d\delta}=\dfrac{a_1 H^{1 + 2 \alpha}}{\delta^2} -\dfrac{ H (2 + \alpha)}{(3 + 2 \alpha) }\delta ^{2\alpha-2}+\mathcal{O}( \delta ^0).\label{a20}
\end{align}
Integrating with respect to $ \delta $ we can find the entanglement entropy for kink in a deformed CFT as 
\begin{align}
S\vert _{k}=S_{(0,k)} +S_{(1,k)},
\end{align}
where
\begin{align}
S_{(0,k)}=-\frac{4\pi L^2}{l_p ^2}\left[\log\left(\frac{\delta}{H}\right)\left(\int _{h_0}^{0} dh\big(\dfrac{\sqrt{1+h^2+\dot{h}^2}}{ \dot{h}h^2}+\frac{1}{h^2})-\frac{1}{h_0}\right)-\dfrac{H}{\delta}\right]
\end{align}
is the he entanglement entropy of kink for the pure AdS case, consistent with the existing results \cite{Myers:2012vs}, and
\begin{equation}
S_{(1,k)}=\left\{ \begin{array}{ll}
	\frac{4\pi L^2}{l_p ^2}\dfrac{\mu^{2\alpha} H\delta ^{2\alpha-1}}{2(2 \alpha-1)} \quad & 0 <\alpha <\frac{1}{2}\nonumber\\
\frac{4\pi L^2}{l_p ^2}\dfrac{\mu^{2\alpha} H}{2} \log\left(\mu\delta\right) \quad &\alpha=\frac{1}{2}\nonumber\\
0\quad &  \alpha>\frac{1}{2},\alpha\neq\frac{3}{2} \end{array} \right.
\end{equation}
is the first order correction of the entanglement entropy due to deformation with a relevant operator. This result also is consistent with Ref. \cite{prvizi} (see table~\ref{tab1}). The important point is that here we have a new universal logarithmic correction for the case $\alpha=1/2$.  Considering the definition of $\alpha$ in the previous section, this shows that for a special tuning $(ML)^2=-5/2$ we have new logarithmic term in entanglement entropy. On the other hand we have no new UV correction for $\alpha >1/2$.

\subsection{cone $ c_n $}
\label{cone}
In this subsection, we give some details for the case $n=1$ corresponding to $d=4$, while for $n=2$ and $n=3$ we give only the results. The entanglement entropy of the cone $c_1$ is achieved
by minimizing the RT surface which may be formulated by  $ \rho =\rho(z, \theta) $ as follows
\begin{align}
S\vert_{c_1}=\frac{2\pi \Omega _1 L^3}{l_{p }^3}\int  d z d\theta\dfrac{\rho\sin(\theta)\sqrt{\frac{\dot{\rho}^2+\rho ^2}{f(z)}+{\rho '}^2 \rho^2 }}
{z^3},\label{b3}
\end{align}
where $ \dot{\rho}=\partial _{\theta} \rho $ and $ \rho '=\partial _{z} \rho $. For this reason we should solve the following equation of motion
\begin{align}
-6 z \sin(\theta) \rho \dot{\rho}^2
+ 2 z \cos(\theta) \dot{\rho}^3
 - 4 z \sin(\theta) \rho^3 (1 + f{ \rho '}^2)+  \sin(\theta) \rho^4 \big(z f' \rho' 
- 6 f^2 {\rho '}^3 + f (-6 \rho '  &\nonumber\\ +2 z \rho '')\big)
+\rho ^2 \big(2 z \sin(\theta)\ddot{\rho} (1 + f {\rho ' }^2)+2 z \dot{\rho}
\left(\cos(\theta) + f \rho ' (\cos(\theta) \rho ' 
 - 2 \sin(\theta)\dot{\rho  }')\right) & \nonumber\\
 +\sin(\theta)\dot{\rho}^2\left(z f' \rho ' 
 +f(-6 \rho '+ 2 z\rho'')\right)\big)& =0.
\label{b4}
\end{align}
Using the general ensatz \eqref{a5}, we can find iterative equations for the unknown functions $h(\theta)$, $g_2(\theta)$. So the entropy functional $S\vert_{c_1}$ would be written peturbatively as
\begin{align}
S\vert_{c_1}=&-\frac{4\pi ^2 L^3}{l_p ^3}(I_0 +\mu^{2\alpha} I_1+\cdots),\label{b6}
\end{align}
where
\begin{align}
&I_k=\int_{z_m} ^{\delta }dz z^{2 \alpha k-1} \int_ {h_0}^{h_{1c}}dh G_k(h),\label{bb6}
\end{align}
in which
\begin{align}
&G_0(h)=\dfrac{\sin(\theta)\sqrt{1+h^2 +\dot{h}^2}}{\dot{h}h^3},\label{d4}\\
&G_1(h)=\dfrac{\sin(\theta)\left(-\dot{h}^2 -h^2(1+2 \dot{h}\dot{g}_2)+4 h^3 g_2 +2 h (3 +2 \alpha +\dot{h}^2) g_2\right)}{ 2 h^3 \dot{h}\sqrt{1+h^2 +\dot{h}^2}}, \label{d5}
\end{align}
where as before we have changed the variable $\theta$ to $h$.
Now let us insert the ensatz \eqref{a5} in the equation of motion \eqref{b4}. Assuming $ y =\sin(\theta) $, we have $ \dot{h}(y) = \sqrt{1-y^2}/y' $ and $ \ddot{h}(y) = -((1 - y^2)y''+
y{y'}^2)/{y'}^3 $ where $ y' = dy/dh $ and $y''=d^2y/dh^2$. Hence the zeroth order part of the equation of motion reads
\begin{align}
&(3+h^2) y (-1+y^2) y'+h(1+h^2)(-1+2 y^2){ y'}^2-(3 +5 h^2 +2 h^4) y {y'}^3 \nonumber\\
&-h (-1 +y^2)\left(-1 +y^2+(1+h^2) y y'' \right)=0.
\label{b7}
\end{align}
In the next order we can use $\dot{g_2}=g_2' \dot{h}$ and $\ddot{g_2}=g_2''\dot{h}^2 +g_2' \ddot{h}$, and relations of $\dot{h}$ and $\ddot{h}$ in terms of $y$, $y'$ and $y''$, to find the following equation
\begin{align}
&h g_2 \big(-2 h (-1 + y^2)^2 (2 + h^2 + 2 \alpha) + 
    y (-1 + y^2) \left(5 + 3 h^4 + 12 \alpha+ 4 \alpha^2 + 
       4 h^2 (1 + \alpha + \alpha^2)\right)\nonumber\\& \times y' +(1 + h^2)  y
      \left(9 + 2 h^4 + 6 \alpha + 
       h^2 (11 + 4 \alpha - 4 \alpha^2)\right) {y'}^3\big) - 
 h (-1 + y^2)^2 \left(1 + 
    2 (h^2 + h^4) g_2'\right) \nonumber\\& + (1 + 
    h^2)y
     \times {y'}^3\left(3 - h^2 (-3 + \alpha) + 
    h^2 (3 + 5 h^2 + 2 h^4) g_2 '\right) + 
 y (-1 + y^2) y'\big(-2 h^2 + \alpha + h^2 \nonumber\\& \times \alpha+ 
    h^2 (1 + h^2) (7 + 5 h^2 + 4 \alpha) 
   g_2'+ h^3 (1 + h^2)^2 g_2''\big)=0.
\label{b8}
\end{align}
 The Eq. \eqref{b7} should be solved with the initial conditions  $ y = sin(\Omega) $ at $ h = 0 $ and $y'=0$ at $h=h_0$ where $h_0=h(\theta=0)$ is the value of $h$ at the turning point. This equation can not be solved exactly. However, we are only interested in the behavior of the quantities near the boundary $h=0$. Hence, we only need to know the expansion of $y(h)$ around $h=0$ which satisfy Eq. \eqref{b7} and the boundary conditions. The result turns out to be as follows 
\begin{align}
y =\sin(\Omega) - \frac{1}{4} \cos(\Omega) \cot(\Omega) h^2 +\left(\frac{1}{64} (3 - \cos(2\Omega)) \cot(\Omega)^2 \csc(\Omega) \log(h)\right) h^4+\mathcal{O}(h^4).\label{b9}
\end{align}
 Now we want to find the expansion of $g_2(h)$ near the boundary. Since $\rho$ should be finite in the limit $h \rightarrow 0$ and $z \rightarrow 0$, the most sigular term in the expansion of $g_2$ should bo of order $h^{1-2\alpha}$. Inserting a power expansion for $g_2$ in terms of $h$ which begins from $h^{-5}$ into Eq. (\ref{b8}) gives the following result
\begin{align}
g_2=\dfrac{a_1}{h^{2\alpha -1}}+\dfrac{a_2}{h^{2\alpha -3}}+\frac{a_3}{ h}+a_4 h+\left(\dfrac{b_1}{h^{2\alpha -3}}+b_2 h\right)\log(h)+\mathcal{O}(h^3),
\label{b10}
\end{align}
where
\begin{align}
&a_2=\frac{a_1}{96} \csc(2\Omega)^2\left(39 + 48 \alpha- 16 \alpha ^2 + 76 \cos(2\Omega) + (37 - 48 \alpha + 16 \alpha ^2)\cos(4 \Omega) \right)\\
&a_3=-\dfrac{-2 + \alpha}{4 (-1 + \alpha^2)}\\
&b_1=-\frac{a_1}{8}\left(-3 + \cos(2 \Omega)\right) \csc(\Omega)^2\\
&b_2=\dfrac{3\left(5 + 4 \cos(2 \Omega) \right) \csc(2 \Omega)^2}{32 (2 + \alpha) (-1 + \alpha ^2)}
\end{align}
The undetermined constant ( $a_1$) in Eq. (\ref{b10}) should be fixed from the boundary conditions  \eqref{d2} and  \eqref{d3}. 
Finally we should insert our results concerning $\sin \theta$, $\dot{h}$ and $g_2$  in terms of $h$ in Eqs. \eqref{d4} and  \eqref{d5} for $G_0$ and $G_1$  to find their needed singular terms as
\begin{align}
&G_0(h) \sim -\dfrac{\sin(\Omega)}{h^3}+\dfrac{\cos(\Omega)\cot(\Omega)}{8 h}+\mathcal{O}\left(h\right), \label{n3}\\
&G_1(h)\sim -\dfrac{2a_1\alpha \sin(\Omega)}{h^{2\alpha +1}}+\dfrac{(-3 + \alpha+ \alpha ^2) \sin(\Omega)}{2(-1 + \alpha^2)h^3}+\mathcal{O}\left(h^{-1}\right) .
 \label{n4}
\end{align}

Now, we can use Eqs.  \eqref{n3} and  \eqref{n4} in the integrands of Eq.  \eqref{bb6} to find their behavior near the boundary: 
\begin{align}
I_0=&I_0'+I_0''\nonumber\\
&\int_{z_m} ^{\delta}\frac{dz}{z}  \int _{h_0}^{h_{1c}} dh\left(G_0(h)+\dfrac{\sin(\Omega)}{h^3}-\dfrac{\cos(\Omega)\cot(\Omega)}{8 h}\right)\nonumber\\
&+\int_{z_m} ^{\delta}\frac{dz}{z}  \int _{h_0}^{h_{1c}} dh\left(-\dfrac{\sin(\Omega)}{h^3}+\dfrac{\cos(\Omega)\cot(\Omega)}{8 h}\right), \label{n6}
\end{align}
\begin{align}
I_1=&I_1'+I_1''\nonumber\\
&\int_{z_m} ^{\delta}z^{2\alpha- 1} dz\int _{h_0}^{h_{1c}} dh\left(G_1(h)+\dfrac{2a_1\alpha \sin(\Omega)}{h^{2\alpha +1}}-\dfrac{(-3 + \alpha+ \alpha ^2) \sin(\Omega)}{2(-1 + \alpha^2)h^3}\right)\nonumber\\
&-\int_{z_m} ^{\delta} z^{2\alpha -1} dz\int _{h_0}^{h_{1c}}dh \left(\dfrac{2a_1\alpha \sin(\Omega)}{h^{2\alpha +1}}-\dfrac{(-3 + \alpha+ \alpha ^2) \sin(\Omega)}{2(-1 + \alpha^2)h^3}\right), \label{n7}
\end{align}
In order to find the singular terms of the integrals  \eqref{n6} and  \eqref{n7} we should find $h_{1c}$ similar to what did in Eq. \eqref{a155}. The result is 
\begin{align}
h_{1c}(\delta)=&\frac{1}{H}\delta +a_1\mu^{2\alpha} H^{2\alpha -3}\delta ^3+a_2 \mu^{2\alpha} H^{2\alpha-5}\delta ^5+\mu^{2\alpha}\frac{a_3}{H}\delta ^{2\alpha+1}+\mu^{2\alpha} \frac{a_4}{H^3}\delta ^{2\alpha+3}\nonumber\\
&+\left(b_1 \mu^{2\alpha} H^{2\alpha-5}\delta ^5+\mu^{2\alpha} \frac{b_2}{H^3}\delta ^{2\alpha+3}\right)\log\delta+\mathcal{O}(\delta ^{2\alpha+5}).\label{b'11}
\end{align}
Following the same technique for calculating derivatives of the singular terms in terms of the cut-off parameter and integrating it again, we find the entanglement entropy of cone as follows

\begin{align}
S\vert_{c_1}=S\vert_{(0,c_1)}+S\vert_{(1,c_1)}
\end{align}
\begin{align}
&S\vert_{(0,c_1)}=-\frac{4\pi ^2 L^3}{l_p ^3}\left(-\dfrac{\sin(\Omega) H^2}{4\delta ^2}+\dfrac{\cos(\Omega)\cot(\Omega)}{16}{\log ^2\left(\frac{\delta}{H}\right)}+q_1\log\left(\frac{\delta}{H}\right)\right),\\
&S\vert_{(1,c_1)}=\left\{ \begin{array}{ll}
	-\frac{4\pi ^2 L^3}{l_p ^3}\left(\dfrac{H^2\mu^{2\alpha}  \sin(\Omega)}{8 (1 - \alpha)\delta ^{2 - 2 \alpha}}\right), \quad & 0 <\alpha <1\nonumber\\
-\frac{4\pi ^2 L^3}{l_p ^3}\left(-\dfrac{H^2 \mu^{2\alpha} \sin(\Omega)}{4} \log (\mu\delta) \right), \quad &\alpha=1 \nonumber\\
0\quad &  \alpha> 1 \end{array} \right.
\end{align}

\begin{table}[tbp]
\centering	
\begin{tabular}{|c|c|c|c|c|c|}
			\hline
			$ d$ &Geometry  &Natural & Geometrical& $ \alpha$& New \\
			&   &  &  singulrities & & divergences for EE \\
			\hline
			&   &&&$ 0<\alpha<\frac{1}{2}  $ &  $1/\delta ^{1-2\alpha}$ \\
			3&$ k$ & $1/\delta $&$\log \delta$&$\alpha =\frac{1}{2}$& $\log\delta $\\
			&&&&$\alpha >\frac{1}{2}$& 0 \\
			\hline
			\hline
			& & &&$ 0<\alpha<1  $ &  $1/\delta ^{2-2\alpha}$ \\
			4&$ c_1$ &$1/\delta^2,\log\delta$&$\log^2\delta$ &$\alpha =1$& $\log\delta $\\
			&&&&$\alpha >1$& 0\\
			\hline
			\hline
			& &&& $ 0<\alpha<3/2,\alpha\neq1/2  $ &  $1/\delta ^{1-2\alpha}$,$1/\delta ^{3-2\alpha}$ \\
			5&$ c_2$&$1/\delta,1/\delta^3$&  $\log\delta $& $\alpha =1/2$ & $1/\delta^2$, $\log\delta $\\
			&& && $\alpha =3/2$ & $\log\delta $\\
			&&&&$\alpha >3/2$& 0\\
			\hline
			\hline
			& & &&$ 0<\alpha<2,\alpha\neq1  $ &  $1/\delta ^{2-2\alpha}$,$1/\delta ^{4-2\alpha}$ \\
			6&$ c_3$ &$1/\delta^2,1/\delta^4,\log\delta$&  $\log^2\delta $& $\alpha =1$ & $1/\delta^2$, $\log\delta $\\
			&& && $\alpha =2$ & $\log\delta $\\
			&&&&$\alpha >2$& 0\\
			\hline
\end{tabular}
\caption{\label{tab1}Entanglement entropy for kink-cones}
\end{table}
where
\begin{align}
&q_1=-\dfrac{\cos(\Omega)\cot(\Omega)}{8}\log(h_0)-\dfrac{\sin(\Omega)}{2 h_0^2}+ \int _{h_0}^{0} dh\bigg(\dfrac{\sin(\theta)\sqrt{1+h^2+\dot{h}^2}}{ \dot{h}h^3} +\dfrac{\sin(\Omega)}{h^3}\nonumber\\
&-\dfrac{\cos(\Omega)\cot(\Omega)}{8 h}\bigg) .
\end{align}

The above calculation can also be done for higher dimensional cones, i.e. for $d=5$ and $d=6$ corresponding to $n=2$ and $n=3$ in Eq. \eqref{a0}, respectively (see the appendix~\ref{Entropy1} for our final results).  In table~\ref{tab1} above, we have shown the nature of new singularities which emerge due to deformation of theory by a relevant operator. These singularities should be accompanied by the ordinary UV singularities, as well as singularities due to needle  points of the subregions considered. As is seen, for the case $c_1$ with the special tuning $M^2L^2=-3$ we have new logarithmic corrections in entanglement entropy. The same thing happens for example for $c_2$ with special tunings $M^2L^2=-9/4$ and $-21/4$.

\subsection{crease $ k\times R^l $}
\label{crease1}
Consider the case $n=0$ for arbitrary $l$  in the general metric (\ref{a0}),  where our singular subregion is limited to the region
 $ \theta \in [-\Omega, \Omega] $, $ \rho \in  [0,H] $  and $ x_i \in [-\frac{\tilde{H}}{2},\frac{\tilde{H}}{2}] $ where $H$ and $ \tilde{H} $ are IR cut-off. Assuming $ \rho =\rho (z ,\theta) $ as the RT surface, we need to optimize the following integral
\begin{align}
S\vert_{k \times R}=\frac{2\pi L^3\tilde{H}}{l_p ^3}\int d z d\theta \dfrac{\sqrt{\frac{\dot{\rho}^2+\rho ^2}{f(z)}+{\rho '}^2 \rho^2 }}
{z^3}.\label{c3}
\end{align} 
Using the general ensatz \eqref{a5} and  inserting it in the equation  optimizing the integral (\ref{c3}), we can expand the expression of the  entanglement entropy as
\begin{align}
S\vert_{k\times R}=-\frac{4\pi L^3\tilde{H}}{l_p ^3}(I_1 +\mu^{2\alpha} I_2 + \cdots),\label{c6}
\end{align}
where
\begin{align}
I_1=\int_{z_m} ^{\delta }\frac{dz}{z^2} \int_ {h_0}^{h_{1c}} \frac{dh}{\dot{h}}\dfrac{\sqrt{1+h^2 +\dot{h}^2}}{ h^2},\label{s4}
\end{align}
\begin{align}
I_2=\int_{z_m} ^{\delta }dz z^{2\alpha -2} \int_ {h_0}^{h_{1c}} \frac{dh}{\dot{h}}\dfrac{ (-\dot{h}^2-h^2(1+2 \dot{h}\dot{g}_2)+4 (1+\alpha) h g_2+2 h^3 g_2)}{2 h^2 \sqrt{1+h^2+\dot{h}^2}}.\label{s5}
\end{align}

Similar to our treatment for kink in section 2, we find in zeroth order the following constant of motion (see Eq.  \eqref{a10})
\begin{align}
K_4=\dfrac{(1+h^2)^{\frac{3}{2}}}{h^3\sqrt{1+h^2+\dot{h}^2}}.\label{c5}
\end{align}
In the first order with respect to $\mu$ the singular behavior of $g_2$ near the boundary turns out to be given (similar to Eq. \eqref{a13}) by the following expansion with respect to the variable $h$
\begin{align}
g_2 =\dfrac{a_1}{h^{2\alpha +1}}+\frac{a_2}{ h^{2\alpha-1}}+\frac{a_3}{ h^{2\alpha-3}}+\frac{a_4}{ h}+a_5 h+\mathcal{O}(h^{3}).\label{s1}
\end{align}
where
\begin{align}
&a_2=-\frac{a_1}{6} (-3 - 7 \alpha + 2 \alpha^2),\label{h1}\\
&a_3=\frac{a_1}{96} (-18 - 3 \alpha+ 71 \alpha^2 - 36 \alpha^3 + 
   4 \alpha^4),\label{h2}\\
&a_4=-\dfrac{1}{4(2 + \alpha)},\label{h3}\\
&a_5=\dfrac{9}{8 (2 + \alpha) (3 + 4 \alpha+ \alpha^2)}\label{h5}.
\end{align}
The quantity $h_{1c} = h(\Omega-\epsilon )$ turns out to have the following expansion (similar to Eq. \eqref{a155}) with respect to the UV cut-off
\begin{align}
h_{1c}(\delta)=&\left(\frac{1}{H}+a_1 \mu^{2\alpha} H^{2\alpha -1}\right)\delta +a_2 \mu^{2\alpha} H^{2\alpha -3}\delta ^3+a_3 \mu^{2\alpha} H^{2\alpha-5}\delta ^5\nonumber\\&+ \mu^{2\alpha}\frac{a_4}{H}\delta ^{2\alpha+1}+\mu^{2\alpha} \frac{a_5}{H^3}\delta ^{2\alpha+3}+
\mathcal{O}(\delta ^{2\alpha+5}).\label{ac10}
\end{align}
Using Eqs.  \eqref{c5},  \eqref{s1} and  \eqref{ac10} we can expand the integrand of Eqs. \eqref{s4} and \eqref{s5}. Then the singular terms with respect to the UV cut-off can be found similar to Eqs \eqref{aa15}--\eqref{a20}.
\begin{align}
S_{k\times R}=S_{(0,k\times R)}+S_{(1,k\times R)},
\end{align}

\begin{align}
S_{(0,k\times R)}=-\frac{4\pi L^3\tilde{H}}{l_p ^3}\left[-\frac{1}{\delta} \int_ {h_0}^{0} dh\left(\dfrac{\sqrt{1+h^2 +\dot{h}^2}}{\dot{h} h^2}+\frac{1}{h^2}\right)-\frac{H}{2\delta ^2}+\frac{1}{h_0 \delta}\right],
\label{c8}
\end{align}
\begin{align}
&S_{(1,k\times R),0<\alpha<1,\alpha\neq\frac{1}{2}}=\nonumber\\
&-\frac{4\pi L^3\tilde{H}}{l_p ^3}\mu^{2\alpha}\bigg[\frac{1}{(2\alpha-1)\delta ^{1-2\alpha}}\int_ {h_0}^{0}dh \bigg(\dfrac{ (-\dot{h}^2-h^2(1+2 \dot{h}\dot{g}_2)+4 (1+\alpha) h g_2+2 h^3 g_2)}{2 h^2\dot{h} \sqrt{1+h^2+\dot{h}^2}}\nonumber\\
&-\bigg(a_5 +  \dfrac{(1 - 2 a_4 )}{2 h^2} + 
 h^{-2 \alpha} \big(a_2  (1 - 2 \alpha)  -\dfrac{a_1 (1 + 2 \alpha)}{h^2}\big)\bigg) \bigg)\nonumber\\
&-\dfrac{h_0^{-1-2\alpha}\left (2 a_2 h_0^2 + (-1 + 2 a_4) h_0^{2 \alpha} +   2 a_5 h_0^{2 + 2 \alpha} + 2 a_1\right) }{2(2\alpha-1)\delta ^{1-2\alpha }}- \dfrac{H}{4(\alpha -1)\delta ^{2-2\alpha}}\bigg],
\end{align}
\begin{align}
&S_{(1,k\times R),\alpha=\frac{1}{2}}=\nonumber\\&
\frac{4\pi L^3\tilde{H}\mu^{2\alpha}}{l_p ^3}\bigg[\dfrac{H}{2 \delta} -\dfrac{(-21 h_0 + 3 h_0^3 + 35 a_1 + 35 h_0^2 a_1}{35 h_0^2} \log(\mu\delta)+\nonumber\\
 &+\log (\mu\delta)\int_ {h_0}^{0}dh \bigg(\dfrac{ -\dot{h}^2-h^2(1+2 \dot{h}\dot{g}_2)+6 h g_2+2 h^3 g_2}{2 \dot{h}h^2 \sqrt{1+h^2+\dot{h}^2}}-\big(\dfrac{21 h + 3 h^3 - 70 a_1}{35 h^3}\big) \bigg)\bigg],\\
&S_{(1,k\times R),\alpha=1}=\frac{4\pi L^3\tilde{H}\mu^{2\alpha}}{l_p ^3}\dfrac{  H}{2}\log(\mu \delta), \\ & S_{(1,k\times R), \alpha>1}=0.
\end{align}
 The above calculation can also be done for higher dimensional cresaes, i.e. for $d=4$, $d=5$ and $d=6$ corresponding to $m=1$, $m=2$ and $m=3$ in Eq. \eqref{a0} respectively (see the appendix~\ref{Entropy1}  for our final results). In table~\ref{tab2} below, we have shown the nature of new singularities which emerge due to deformation of theory by a relevant operator.
\begin{table}[tbp]
\centering
\begin{tabular}{|c|c|c|c|c|c|}
\hline
$ d$ &Geometry & Natural & Geometrical& $ \alpha$ & New \\
 &   &   &singularities&&  divergences for EE \\
\hline
 &   &&& $ 0<\alpha<1,\alpha\neq\frac{1}{2} $&  $1/\delta ^{1-2\alpha}$, $1/\delta ^{2-2\alpha}$ \\
4&$ k\times R$ &$1/\delta ^2$&$1/\delta $& $\alpha =\frac{1}{2}$& $\log\delta $, $1/\delta$ \\
& &&& $\alpha =1$& $\log\delta $ \\
&&&&$\alpha >1$& 0 \\
\hline
 \hline
 &   & &&$ 0<\alpha<3/2,\alpha\neq1  $ &  $1/\delta ^{2-2\alpha}$, $1/\delta ^{3-2\alpha}$ \\
5&$ k\times R^2$&$1/\delta ^3$&$1/\delta ^2$ & $\alpha =1$& $\log\delta $, $1/\delta$ \\
& & &&$\alpha =\frac{3}{2}$& $\log\delta $ \\
&&&&$\alpha >\dfrac{3}{2}$& 0 \\
\hline
 \hline
 &   &&& $ 0<\alpha<2,\alpha\neq\frac{3}{2}  $ &  $1/\delta ^{3-2\alpha}$, $1/\delta ^{4-2\alpha}$ \\
6&$ k\times R^3$&$1/\delta ^4$&$1/\delta^3 $ & $\alpha =\frac{3}{2}$& $\log\delta $, $1/\delta$ \\
& &&& $\alpha =2$& $\log\delta $ \\
&&&&$\alpha >2$& 0 \\
\hline
\end{tabular}
\caption{\label{tab2}Entanglement entropy for $ k\times R^l $}
\end{table}

\subsection{crease $ c_n\times R^l $}
\label{crease2}
Now we consider some examples of the more general form of the bulk geometry introduced earlier in equation \eqref{a0}. For  $ l = 1 $ and $ n=1 $ the singular subregion in the boundary is defined as
$ \theta \in \left[ 0; \Omega \right] $, $ \phi \in \left[ 0; 2 \pi \right] $, $ \rho \in\left[ 0;  H \right] $ and $ x^l \in \left[ \frac{\tilde{H}}{2};- \frac{\tilde{H}}{2} \right] $. The entanglement entropy is obtained by minimizing the area of RT surface given by
\begin{align}
S\vert_{c_1\times R}=\frac{2\pi \Omega _1 L^4\tilde{H}}{l_{p }^4}\int  d z d\theta\dfrac{\rho\sin(\theta)\sqrt{\frac{\dot{\rho}^2+\rho ^2}{f(z)}+{\rho '}^2 \rho^2 }}
{z^4},\label{c10}
\end{align}
where $ \rho =\rho (z ,\theta) $ gives the extension of subregion to the bulk. As before this quantity can be expanded as given by the ensatz \eqref{a5}. Similar to case of $ c_n $ we find the expansions of the quantities $y=\sin \theta$ and $g_2$ near the boundary (i.e. in the limit $h\rightarrow 0$) as follows
\begin{align}
&y=\sin(\Omega) - \frac{1}{6}\cos(\Omega)\cot(\Omega) h^2 +  \frac{1}{432}\left(-19 + 5 \cos(2\Omega)\right) \cot^2(\Omega)\csc(\Omega) h^4+\mathcal{O}(h^5),\\
&g_2(h)=\frac{a_1}{h^{2 \alpha-2}}+\frac{a_2}{h^{2\alpha-4}}+\dfrac{a_3}{h} +a_4 h +\mathcal{O}(h^3).
\end{align}
where
\begin{align}
&a_2=\frac{1}{252} a_1 \left(-235 + 144 \alpha- 
   36 \alpha^2 + (179 - 144 \alpha + 36 \alpha^2) \cos(2 \Omega) \right)\csc^2(\Omega),\\
&a_3=\dfrac{3-\alpha}{-6 - 2 \alpha+ 4 \alpha^2},\\
&a_4=\dfrac{\alpha \left(-14 + \alpha + (4 + \alpha) \cos(2 \Omega)\right) \csc^2(\Omega)}{9 (6 - 7 \alpha- 13 \alpha^2 + 4 \alpha^3 + 4 \alpha^4)}.
\end{align}
We also need to know the limiting value $h_{1c}$ in terms of the UV-cutoff. The result is
\begin{align}
h_{1c}(\delta)=\frac{\delta}{H} +a_1 \mu ^{2\alpha} H^{2\alpha -4}\delta ^4+a_2 \mu ^{2\alpha} H^{2\alpha-6}\delta ^6+ \mu ^{2\alpha}\frac{a_3}{H}\delta ^{2\alpha+1}+\mu ^{2\alpha} \frac{a_4}{H^3}\delta ^{2\alpha+3}+\mathcal{O}(\delta ^{2\alpha+5}).\label{a15}
\end{align}
Inserting all the data in the integral \eqref{c10} leads to the following result
\begin{align}
S_{c_1\times R}=S_{(0,c_1\times R)}+S_{(1,c_1\times R)},
\end{align}
where
\begin{align}
&S_{(1,c_1\times R),0<\alpha<\frac{3}{2},\alpha\neq\frac{1}{2}}=\nonumber \\&-\frac{4\pi ^2 L^4\tilde{H}\mu ^{2\alpha} }{l_p ^4}\bigg[\dfrac{H^2\sin(\Omega)}{(12 - 8 \alpha) \delta ^{3-2\alpha}}-\dfrac{h0^{-2 (1 + \alpha)}}{ \delta^{1-2\alpha}36 (1 - 
     2 \alpha)^2 (1 + \alpha) (2 + \alpha) (-3 + 
     2 \alpha)}\nonumber\\
 &\times\bigg(162 h_0^{2 \alpha} - 216 h_0^3 a_1 - 261 h_0^{2 \alpha} \alpha+ 
    16 h_0^{2 + 2 \alpha} \alpha+ 252 h_0^3 a_1 \alpha- 
    171 h_0^{2 \alpha} \alpha^2 + 
    4 h_0^{2 + 2 \alpha} \alpha^2 \nonumber\\
    &+ 468 h_0^3 a_1 \alpha^2 + 
    72 h_0^{2 \alpha} \alpha^3 - 144 h_0^3 a_1\alpha^3 + 
    36 h_0^{2 \alpha} \alpha^4 - 144 h_0^3 a_1 \alpha^4
     -  4 h_0^{2 + 2 \alpha} (-14 + \alpha) \alpha \times \nonumber\\
     &\csc^2(\Omega)+ 
    2 h_0^{2 + 2 \alpha}
     \cot^2(\Omega)\big(-24 + 4 \alpha + 
       10 \alpha ^2
       + \left(18 - 37 \alpha - 7 \alpha^2 + 
          16 \alpha^3 + 4 \alpha^4\right) \log(
         h_0) \nonumber\\&+ \left(-18 + 37 \alpha+ 7 \alpha^2 - 16 \alpha^3 - 
          4 \alpha^4\right) \log(\delta/H)\big) \bigg)\sin(\Omega)\nonumber\\
       &+ \frac{1}{\delta ^{1-2\alpha}(-1 + 2 \alpha)}\int_{h_0} ^0 dh\dfrac{\sin(\theta) \left(-
\dot{h}^2 - h^2 (1 + 2 \dot{h}\dot{g_2}) +4 h^3 g_2 +  2 h (3 + 2 \alpha+ \dot{h}^2)g_2\right)}{2h^3\dot{h}\sqrt{1+h^2 +\dot{h}^2}}\nonumber\\
&-\big(-\dfrac{ (-9 + 5 \alpha+ 2 \alpha^2) \cos(\Omega) \cot(\Omega)}{18 h (-3 - \alpha + 2 \alpha^2)}+h^{-2 \alpha} a_1 (1 - 2 \alpha) \sin(\Omega)\nonumber\\
&+\dfrac{ (-9 + \alpha + 2 \alpha^2) \sin(\Omega)}{ h^3 (-6 - 2 \alpha + 4 \alpha^2)} \big)\bigg],
\end{align}

\begin{align}
&S_{(1,c_1\times R),\alpha=\frac{1}{2}}=\nonumber \\&-\frac{4\pi ^2 L^4\tilde{H}\mu ^{2\alpha} }{l_p ^4}\Bigg[\dfrac{H^2 \sin(\Omega)}{8\delta ^2}+  \frac{1}{72}\log(\mu\delta)\bigg(-\frac{10}{3} \cos(\Omega) \cot(\Omega) +    8 \cos(\Omega) \cot(\Omega) \nonumber\\ &\times\log(h_0)\- 
    4 \cos(\Omega) \cot(\Omega)\log(\mu\delta)+ 
    9 \left(\frac{16}{3 h_0^2} - 8 a_1\right)  \sin(\Omega)\bigg)\nonumber\\
   +& \log(\mu \delta )\int_{h_0} ^0 dh\bigg(\dfrac{\sin(\theta) \left(-
\dot{h}^2 - h^2 (1 + 2 \dot{h}\dot{g_2}) +4 h^3 g_2 +  2 h (4+ \dot{h}^2)g_2\right)}{2h^3\dot{h}\sqrt{1+h^2 +\dot{h}^2}}\nonumber\\
&+\dfrac{h^2 \cos(\Omega) \cot(\Omega) - 12 \sin(\Omega)}{9 h^3}\bigg)\Bigg],\\
&S_{(1,c_1\times R),\alpha=\frac{3}{2}}=-\frac{4\pi ^2 L^4\tilde{H}\mu ^{2\alpha} }{l_p ^4}\left(-\frac{1}{4} H^2  \log(\mu\delta)\sin(\Omega)\right), \\& S_{(1,c_1\times R),\alpha>\frac{3}{2}}=0.
\end{align}
For crease $ c_1 \times R^2$ the final results are given in appendix~\ref{Entropy1}. Table \ref{tab3} below, shows the nature of new singular terms similar to previous cases. 
\begin{table}[tbp]
\centering
\begin{tabular}{|c|c|c|c|c|c|}
\hline
$ d$ &Geometry &Natural & Geometrical& $ \alpha$ & New \\
 &   &  & singularities && divergences for EE \\
\hline
 &   & &&$ 0<\alpha<\frac{3}{2},\alpha\neq\frac{1}{2} $& $\log( \delta)/\delta ^{1-2\alpha}$ ,$1/\delta ^{1-2\alpha}$, $1/\delta ^{3-2\alpha}$ \\
5&$ c_1\times R$ &$1/\delta^3$,$1/\delta$&$\log(\delta)/\delta$& $\alpha =\frac{1}{2}$&$\log ^2(\delta) $ ,$\log(\delta) $, $1/\delta^2$ \\
& &&& $\alpha =\frac{3}{2}$& $\log(\delta) $ \\
&&&&$\alpha >\frac{3}{2}$& 0 \\
\hline
 &   &&& $ 0<\alpha<2,\alpha\neq1  $ &$\log\delta/\delta ^{2-2\alpha}$ , $1/\delta ^{2-2\alpha}$, $1/\delta ^{4-2\alpha}$ \\
6&$ c_1\times R^2$& $1/\delta^4$,$1/\delta^2,\log(\delta )$&$\log(\delta)/\delta$& $\alpha =1$& $\log ^2(\delta) $,$\log(\delta) $, $1/\delta^2$ \\
& & &&$\alpha =2$& $\log\delta $ \\
&&&&$\alpha >2$& 0 \\
\hline
 \hline
 &   &&& $ 0<\alpha<2,\alpha\neq\frac{1}{2},1  $ &  $1/\delta ^{1-2\alpha}$,$1/\delta ^{2-2\alpha}$,$1/\delta ^{4-2\alpha}$ \\
6&$ c_2\times R$ &$1/\delta^4$,$1/\delta^2,\log(\delta )$&$1/\delta$& $\alpha =\frac{1}{2}$& $\log(\delta) $, $1/\delta$,$1/\delta^3$ \\
& & &&$\alpha =1$& $\log(\delta) $, $1/\delta^2$\\
& & &&$\alpha =2$& $\log(\delta) $ \\
&&&&$\alpha >2$& 0 \\
\hline
  &   & &&$ 0<\alpha<\frac{5}{2},\alpha\neq1,\frac{3}{2}  $ &$1/\delta ^{2-2\alpha}$  $1/\delta ^{3-2\alpha}$, $1/\delta ^{5-2\alpha}$ \\
7&$ c_2\times R^2$ &$1/\delta^5$,$1/\delta^3,1/\delta$&$1/\delta^2$& $\alpha =1$& $\log(\delta) $, $1/\delta$ ,$1/\delta^3$\\
& & &&$\alpha =\frac{3}{2}$& $\log(\delta) $ ,$1/\delta^2$\\
& & &&$\alpha =\frac{5}{2}$& $\log(\delta) $ \\
&&&&$\alpha >\frac{5}{2}$& 0 \\
\hline
 \hline
\end{tabular}
\caption{\label{tab3}Entanglement entropy for $ c_n\times R^l $}
\end{table}

\section{Subregion Complexity}
\label{Complexity}
{As stated in the introduction, the subregion approach for complexity of a given static state \cite{Alishahiha:2015rta,Ben-Ami:2016qex} in a conformal theory concerns the volume enclosed by the Ryu-Takayanagi surface, i.e.
\begin{align}
\mathcal{C}=\dfrac{V(\gamma)}{8 \pi l G_N},\label{a22}
\end{align}
where $l$ is a characteristic length scale of the bulk geometry, and $ \gamma $ is the RT surface corresponding to the subregion specified in the boundary. Now let us calculate this quantity for the kink given by the metric \eqref{a0}. The volume surrounded by the surface with the induced metric \eqref{ab11} can be given as
\begin{align}
V(\gamma)=&L^3 \int dz d\theta d\rho \dfrac{\rho}{\sqrt{f(z)}z^3}\nonumber\\
&=L^3 \int dz d\theta \dfrac{1}{\sqrt{f(z)}z^3}\int _{\rho(z,\theta)} ^H d\rho \rho\nonumber\\
&=\frac{L^3}{2}\left(-\int dz d\theta \dfrac{1}{\sqrt{f(z)}z^3} \rho ^2+H^2\int dz d\theta \dfrac{1}{\sqrt{f(z)}z^3}\right)\nonumber\\
&\equiv V_1 +V_2,\label{a23}
\end{align}
Using the ansatz (\ref{a5}) in the integrand we find
\begin{align}
V_2=\dfrac{L^3 H^2 \Omega}{2 \delta ^2}+\dfrac{H^2 L^3 \mu ^{2\alpha} \Omega \delta ^ {2\alpha-2}}{4(\alpha-1)}+\mathcal{O}(\delta ^0),
\end{align}
\begin{align}
V_1=&L^3\int_{z_m}^\delta \frac{dz}{z} \int_{h_0}^{h_{1c}} dh \dfrac{1}{h^2 \dot{h}}+L^3\mu ^{2\alpha} \int_{z_m}^\delta dz z^{2\alpha -1}\int _{h_0}^{h_{1c}} dh\dfrac{-1+4 h g_2}{2 h^2\dot{h}}+ \cdots \\. \label{a24} 
\end{align}
Let us search for the divergent terms of $V_1$ using
\begin{align}
\dfrac{1}{h^2 \dot{h}}\sim -k_3 +k_3 h^2+\mathcal{O}(h^3),
\end{align}
\begin{align}
\dfrac{-1+4 h g_2}{2 h^2\dot{h}}\sim -\dfrac{2a_1 k_3}{h^{2 \alpha}} +\frac{5+2\alpha}{2(3+2 \alpha)}k_3+\mathcal{O}(h^{2 -2\alpha}),
\end{align}
where $ a_1 $ is the  coefficient  defined in Eq. \eqref{a13}. In terms of different powers of $\mu$ we have
\begin{align}
V_1=L^3( I_1 +\mu ^{2\alpha} I_2+\cdots),
\end{align}
where
\begin{align}
I_1=\int_{z_m}^\delta \frac{dz}{z} \int_{h_0}^{h_{1c}} dh \dfrac{1}{h^2 \dot{h}},
\end{align}
\begin{align}
I_2=&\int_{z_m}^\delta dz z^{2\alpha -1}\int _{h_0}^{h_{1c}} dh\left(\dfrac{-1+4 h g_2}{2 h^2\dot{h}}+\dfrac{2a_1 k_3}{h^{2 \alpha}} -\frac{5+2\alpha}{2(3+2 \alpha)}k_3\right)\nonumber\\
&-\int_{z_m}^\delta dz z^{2\alpha-1}\int _{h_0}^{h_{1c}} dh\left(+\dfrac{2a_1 k_3}{h^{2 \alpha}} -\frac{5+2\alpha}{2(3+2 \alpha)}k_3\right)\nonumber\\
&=I_1'+I_2'.
\end{align}
In the expansions of the above integrands we encounter terms of the form $\int dz z^n \int dh h^{-m}$ for positive $n$ and $m$. To find the singularities we need just to keep terms with $m \ge n+2$. In the first integral $V_1$, which is just the complexity for the undeformed CFT, the integral over $h$ is finite, so we have a logarithmic divergent term for integration over $z$. However, in the expansions of the terms in the integrals $V_1$ over $h$, we do not find any term for which $m \ge n+2$. The final result reads

\begin{align}
\mathcal{C}_{k}=\mathcal{C}_{(0,k)}+\mathcal{C}{(1,k)}
\end{align}
\begin{align}
\mathcal{C}_{(0,k)}=\frac{L^3}{8 \pi l G}\left(\dfrac{ H^2 \Omega}{2 \delta ^2}+\log(\delta/H)\int_{h_0}^{0} dh \dfrac{1}{h^2 \dot{h}}\right)+finite, 
\end{align}
\begin{align}
\mathcal{C}_{(1,k)}=\left\{ \begin{array}{ll}
	\frac{L^3 \mu ^{2\alpha}}{8 \pi l G}\left(\dfrac{H^2  \Omega}{4(-1+\alpha)\delta ^{2- 2 \alpha}}\right), \quad & 0 <\alpha <1\nonumber\\
\frac{L^3 \mu ^{2\alpha}}{8 \pi l G}\left(\dfrac{H^2  \Omega}{2}\log(\mu \delta) \right), \quad &\alpha=1 \nonumber\\
0 \quad &  \alpha>1\end{array} \right.
\end{align}

We can also find the subregion complexity for all of the cones discussed in the previous section. The final result for $c_1$ is 
\begin{align}
\mathcal{C}_{c_1}=\mathcal{C}_{(0,c_1)}+\mathcal{C}{(1,c_1)}
\end{align}
where
\begin{align}
\mathcal{C}_{(0,c_1)}=\dfrac{L^3}{8 G_N}\left[\dfrac{2\left(1-\cos(\Omega)\right)}{9}\dfrac{H^3}{\delta ^3}-\frac{\cos(\Omega)}{3}\dfrac{ H}{\delta}+\dfrac{\beta(h_0)}{3}\log\left(\dfrac{\delta}{H}\right) \right], 
\end{align}
\begin{table}
\centering
\begin{tabular}{|c|c|c|c|c|c|}
\hline
$ d$ &Geometry  &  Natural&Geometrical& $ \alpha$ & New \\
 &   &   & singularities& & divergences for $\mathcal{C}$ \\
\hline
 &    &&&$0<\alpha<1$ & $1/\delta ^{2-2\alpha}$\\
3&$ k$ &$1/\delta^2$&$\log(\delta) $&$\alpha=1$ &$\log(\delta) $\\
&&&&$\alpha>1$ &0 \\
\hline
 & &&&$ 0<\alpha<\frac{3}{2}  $,$\alpha\neq \frac{1}{2},\frac{3}{2} $ & $1/\delta ^{1-2\alpha}$,$1/\delta ^{3-2\alpha}$\\
4&$ c_1$ &$1/\delta^3$,$1/\delta$&$\log(\delta )$&$\alpha=\frac{1}{2}$ &$\log(\delta) $,$1/\delta^2$\\
&&&&$\alpha=\frac{3}{2}$ &$\log(\delta) $\\
&&&&$\alpha>3/2$ &0 \\
\hline
 & &&&$ 0<\alpha<2  $,$\alpha\neq 1,2 $& $1/\delta ^{2-2\alpha}$,$1/\delta ^{4-2\alpha}$\\
5&$ c_2$ &$1/\delta^4$,$1/\delta^2$,$\log(\delta) $&$\log^2(\delta) $&$\alpha=1$ &$\log(\delta) $,$1/\delta^2$\\
&&&&$\alpha=2$ &$\log(\delta) $\\
&&&&$\alpha>2$ &0 \\
\hline
 & &&&$ 0<\alpha<\frac{5}{2}  $,$\alpha\neq \frac{1}{2},\frac{3}{2},\frac{5}{2} $& $1/\delta ^{1-2\alpha}$,$1/\delta ^{3-2\alpha}$,$1/\delta ^{5-2\alpha}$\\
6&$ c_3$ &$1/\delta^5$,$1/\delta^3$,$1/\delta $&$\log(\delta )$&$\alpha=1/2$ &$\log(\delta) $,$1/\delta^2$,$1/\delta^4$\\
&&&&$\alpha=3/2$ &$\log(\delta) $,$1/\delta^2$\\
&&&&$\alpha=5/2$ &$\log(\delta) $\\
&&&&$\alpha>5/2$ &0 \\
\hline
\end{tabular}
\caption{\label{tab4}Complexity for $k-c_n$}
\end{table}
 \begin{table}
\centering
 \begin{tabular}{|c|c|c|c|c|c|}
\hline
$ d$ &Geometry & Natural&Geometrical& $ \alpha$ & New \\
 &   &  & singularities&& divergences for $\mathcal{C}$\\
\hline
 &   &&& $ 0<\alpha<\frac{3}{2} ,\alpha\neq \frac{1}{2}$&  $1/\delta ^{1-2\alpha}$, $1/\delta ^{3-2\alpha}$ \\
4&$ k\times R$ &$1/\delta$ ,$1/\delta^3$ &-& $\alpha =\frac{1}{2}$& $\log\delta $, $1/\delta^2$ \\
& & &&$\alpha =\frac{3}{2}$& $\log\delta $ \\
&&&&$\alpha >\frac{3}{2}$& 0 \\
\hline
 &   &&& $ 0<\alpha<2,\alpha\neq1  $ &  $1/\delta ^{2-2\alpha}$, $1/\delta ^{4-2\alpha}$ \\
5&$ k\times R^2$&$1/\delta^2$ ,$1/\delta^4$&- & $\alpha =1$& $\log\delta $, $1/\delta^2$ \\
& & &&$\alpha =2$& $\log\delta $ \\
&&&&$\alpha >2$& 0 \\
\hline
 &   &&& $ 0<\alpha<\frac{5}{2},\alpha\neq\frac{3}{2}  $ &  $1/\delta ^{3-2\alpha}$, $1/\delta ^{5-2\alpha}$ \\
6&$ k\times R^3$ &$1/\delta^3$ ,$1/\delta^5$&-& $\alpha =\frac{3}{2}$& $\log\delta $, $1/\delta^2$ \\
& & &&$\alpha =\frac{5}{2}$& $\log\delta $ \\
&&&&$\alpha >\dfrac{5}{2}$& 0 \\
\hline
\end{tabular}
\caption{\label{tab5}Complexity for $ k\times R^l $}
\end{table}
\begin{table}[tbp]
\centering
\begin{tabular}{ |c|c|c|c|c|}
\hline
Geometry & Natural&Geometrical &$ \alpha$ & New \\
   &  & singularities&&  divergences for $\mathcal{C}$ \\
\hline
    &&& $ 0<\alpha<2,\alpha\neq\frac{1}{2},1 $& $1/\delta ^{i-2\alpha}$ ,$i=1,2,4$ \\
$ c_1\times R$ &$\log(\delta) $,$1/\delta^i$,$ i=2,4$& $1/\delta$& $\alpha =\frac{1}{2}$&$\log(\delta) $,$1/\delta $ , $1/\delta^3$ \\
 & &&$\alpha =1$& $\log(\delta) $ , $1/\delta^2$\\
 &&& $\alpha =2$& $\log(\delta) $ \\
&&&$\alpha >2$& 0 \\
\hline
\hline
 &&& $ 0<\alpha<\frac{5}{2},\alpha\neq\frac{1}{2},\frac{3}{2}  $ &  $\log(\delta)/\delta ^{1-2\alpha}$,$1/\delta ^{i-2\alpha}i=1,3,5$ \\
$ c_2\times R$ &$1/\delta^i$,$i=1,3,5$&$\log(\delta)/\delta$& $\alpha =\frac{1}{2}$&  $\log^2(\delta )$,$\log(\delta )$, $1/\delta^2$, $1/\delta^4$ \\
 &&& $\alpha =\frac{3}{2}$& $\log(\delta) $,$1/\delta^2$ \\
 &&& $\alpha =\frac{5}{2}$& $\log(\delta) $\\
&&&$\alpha >\frac{5}{2}$& 0 \\
\hline
 &&  & $ 0<\alpha<3,\alpha\neq1,2  $ &   $\log(\delta)/\delta ^{2-2\alpha}$,$1/\delta ^{i-2\alpha}$, $i=2,4,6$ \\
$ c_2\times R^2$ &$\log(\delta)$,$1/\delta^i$,$ i=2,4,6$&$\log(\delta)/\delta^2$& $\alpha =1$& $\log^2(\delta) $,$\log(\delta) $, $1/\delta^2$, $1/\delta^4$ \\
 &&& $\alpha =2$& $\log(\delta) $, $1/\delta^2$\\
 &&& $\alpha =3$& $\log(\delta) $\\
&&&$\alpha >3$& 0 \\
\hline
\end{tabular}
\caption{\label{tab6} Complexity for $ c_n\times R^l $}
\end{table}
and
\begin{align}
\mathcal{C}_{(1,c_1)}=\left\{ \begin{array}{ll}
	\dfrac{L^3\mu ^{2\alpha}}{8 G_N}\left[-\dfrac{H (-3 + \alpha + \alpha ^2) \cos(\Omega) }{6 (-1 + 2 \alpha) (-1 + \alpha ^2)\delta ^{1-2\alpha}}+\dfrac{ H^3  (-1 + \cos(\Omega))}{\delta^{3 - 2 \alpha}(9-6 \alpha)}\right], \quad & 0 <\alpha <\frac{3}{2},\alpha \neq\frac{1}{2}\nonumber\\
\dfrac{L^3\mu ^{2\alpha}}{8 G_N}\left[\dfrac{ H^3(\cos(\Omega)-1)}{6 \delta^{2}}-\frac{1}{2}H  \cos(\Omega) \log(\mu\delta)\right], \quad &\alpha=\frac{1}{2} \nonumber\\
\dfrac{L^3\mu ^{2\alpha}}{8 G_N}\left[-\frac{1}{3} H^3 (-1 +\cos(\Omega))\log(\mu\delta)\right],\quad &  \alpha= \dfrac{3}{2}\nonumber\\
0, \quad &  \alpha> \dfrac{3}{2}\end{array} \right.
\end{align}

in which
\begin{align}
\beta(h_0)=&2 \int_ {h_0}^{0} dh\bigg(\dfrac{\sin(\theta)}{h^3\dot{h}}+\dfrac{1}{2} \dfrac{\cos( \Omega)}{h^2}-\frac{1}{8}\cot ^2 (\Omega ) \sin (\Omega ) \csc (2\Omega)(3 - \cos (2\Omega)) \log (h) \nonumber\\ &- \dfrac{1}{8}\cos(\Omega)\cot ^2(\Omega)\bigg)
-\dfrac{\cos(\Omega)}{ h_0} -\dfrac{h_0}{4}\cos(\Omega) \cot ^2(\Omega)+\frac{1}{4} h_0( 1-\log(h_0))\cot ^2 (\Omega )  \nonumber\\ &\times\sin (\Omega )
\csc (2\Omega)(3 - \cos (2\Omega)).
\end{align}
In the tables~\ref{tab4},\ref{tab5} and \ref{tab6} we give the nature of UV singular terms for different cases.  

\section{Discussions} 
\label{Discussions}
In this paper we concentrated on the effect of deformation of the boundary CFT due to a relevant  operator achieved by turning on a scalar field in the bulk. The entanglement entropy, as well as the subregion complexity, are important quantities  which demonstrate physical properties of the deformed CFT. The singular behavior of these quantities near the boundary or near the singular point (or wedge) of an assumed singular subregion may be viewed as key points toward investigating the properties of the corresponding CFT. 

Considering the asymptotic  AdS-theory, we introduced a set of subregions characterized by singular points or wedges. For the pure AdS space-time in the bulk (i.e. undeformed theory) two kinds of singularity appears in entanglement entropy and complexity. The first category is the natural  UV singularity of these quantities. However, it is showed that for singular subregions, such as kink, cone or cresae, new kinds of singularity may appear as one  approaches the singular point or wedge. Deformation of the CFT by means of a relevant operator, imposes a third kind of singularity in the final expressions. 

Our  results are summarized in tables  tables~\ref{tab1}--\ref{tab6} in the text. For completeness we have shown the known results for natural as well as geometrical singularities in independent columns (forth and fifth columns). The new kind of singularities due to deformation are $ \mu$-dependent terms which are shown in the sixth column of the corresponding tables. Tables \ref{tab1},\ref{tab2} and \ref{tab3} give the entanglement entropy for kink, cones, and creases $k\times R^l$ and  $C_n\times R^l$, respectively. The tables \ref{tab4},\ref{tab5} and \ref{tab6} give the subregion complexity for the same singular subregions. In the tables we just have indicated the kind of UV divergences. However, in appendices \ref{Entropy1} and \ref{Entropy2} we have given the complete singular terms for entanglement entropy and complexity respectively.

Among different results we point on the following important ones.\\
i) In all cases we have new logarithmic divergent terms in entanglement entropy for particular values of conformal dimension of the relevant operator which corresponds to some special tunings of the parameter $M$ of the scalar field in the bulk.\\ 
ii) The same thing happens for complexity but for different values of conformal dimension. In fact for deformed theory we may have logarithmic corrections in all dimensions for particular values of the conformal dimension.\\
ii) For crease $C_1\times R^l$ we found for entanglement entropy divergent terms of the form $\delta^{-k} \log \delta$ or $(\log \delta)^2$ depending on the values of $l$ and the conformal dimension. The same thing happens for complexity for crease $C_2\times R^l$.

\acknowledgments

 We would like to thank Mohsen Alishahiha, Amin Faraji-Astaneh  and Ali Naseh for fruitful discussions and M. Reza Mohammadi-Mozaffar and Ali Mollabashi for careful reading of the manuscript.

\appendix
\section{Some Results for Entanglement Entropy}
\label{Entropy1}
For the cones $c_n $ (remember $d=n+3$) similar calculations as $c_1$ can be performed. Our final results for $d=5$ read 
\begin{align}
S_{c_2}=S_{(0,c_2)}+S_{(1,c_2)}
\end{align}

\begin{align}
S_{(0,c_2)}=-\frac{8\pi ^2 L^4}{l_p ^4}\left(-\dfrac{\sin ^2(\Omega) H^3}{9 \delta ^3}+\dfrac{4\cos^2(\Omega) H}{9\delta} +q_2\log \big(\frac{\delta}{H}\big)\right),
\label{c8}
\end{align}
\begin{equation}
S_{(1,c_2)}=\left\{ \begin{array}{ll}
-\frac{8\pi ^2 L^4\mu ^{2\alpha}}{l_p ^4}\left(\dfrac{  H^3 \sin ^2(\Omega)}{(18 - 12 \alpha) \delta^{3-2\alpha}}+\dfrac{2 H  (-9 + 5 \alpha+ 2 \alpha^2) \cos^2(\Omega)}{
   9 (1 + \alpha) (3 - 8 \alpha + 4 \alpha ^2) \delta^{1-2\alpha}}\right),\quad &0 <\alpha <\frac{3}{2}, \alpha \neq1 \nonumber\\
-\frac{8\pi ^2 L^4\mu ^{2\alpha}}{l_p ^4}\left(\dfrac{H^3  \sin^2(\Omega)}{12 \delta^2}+\frac{4}{9} H  \cos^2(\Omega)\log \big(\mu\delta\big)\right), \quad &\alpha=1\nonumber\\
-\frac{8\pi ^2 L^4\mu ^{2\alpha}}{l_p ^4}\left(-\frac{1}{6}H^3  \sin^2(\Omega)\log\big(\mu \delta\big)\right), \quad &\alpha=\frac{3}{2} \nonumber\\
0,\quad &  \alpha>\frac{3}{2}. \end{array} \right.
\end{equation}
where
\begin{align}
q_2=&-\dfrac{{\sin(\Omega)}^2}{3 h_0^3}+\dfrac{4 \cos ^2(\Omega)}{9 h_0}+\dfrac{\cos ^2(\Omega) h_0}{9}+ \int _{h_0}^{0} dh \bigg(\dfrac{\sin ^2(\theta)\sqrt{1+h^2+\dot{h}^2}}{ \dot{h}h^4}+\dfrac{\sin ^2(\Omega)}{h^4}\nonumber\\&-\dfrac{4\cos ^2(\Omega)}{9h^2}\bigg).
\end{align}
For cone $c_3$  in d =  6 we find
\begin{align}
S_{c_3}=S_{(0,c_3)}+S_{(1,c_3)}
\end{align}
\begin{align}
S_{(0,c_3)}=&\frac{8\pi ^3 L^5}{l_p ^5}\Bigg[\dfrac{H^4 \sin ^3(\Omega)}{16 \delta ^4}-\dfrac{27 H^2 \cos ^2 (\Omega) \sin(\Omega)}{128 \delta ^2}+q_4 \log\left(\frac{\delta}{H}\right)\nonumber\\
&+\dfrac{9 \cos(\Omega)\cot(\Omega)(31-\cos(2\Omega))}{8192}\log ^ 2 \left(\frac{\delta}{H}\right)\Bigg],
\end{align}
\begin{equation}
S_{(1,c_3)}=\left\{ \begin{array}{ll}
\frac{8\pi ^3 L^5\mu ^{2\alpha}}{l_p ^5}\left(-\dfrac{9 H^2  (-18 + 5 \alpha + 3 \alpha^2) \cos^2(\Omega) \sin(\Omega)}{256 (-2 + \alpha) (-1 + \alpha) (1 + \alpha) \delta ^{2-2\alpha} }-\dfrac{H^4  \sin^3(\Omega)}{(32 - 16 \alpha) \delta^{4-2\alpha}}\right), &0 <\alpha <2, \alpha \neq1 \nonumber\\
\frac{8\pi ^3 L^5\mu ^{2\alpha}}{l_p ^5}\bigg(-\frac{1}{128} H^2 \sin(\Omega)
  \big(45 \cos ^2(\Omega) \log(\mu\delta) +\dfrac{8 H^2 \sin(\Omega)}{\delta ^2}\big) \bigg), \quad &\alpha=1\nonumber\\
\frac{8\pi ^3 L^5\mu ^{2\alpha}}{l_p ^5}\left(\dfrac{H^4  \sin^3(\Omega)}{8}\log (\mu\delta)\right), \quad &\alpha=2\nonumber\\
0,\quad &  \alpha>2. \end{array} \right.
\end{equation}
where
\begin{align}
q_4=&\int_ 0^{ h_0} dh \bigg(\dfrac{\sin ^3(\theta)}{\dot{h} h^5}\sqrt{1+h^2 +\dot{h}^2}+\dfrac{\sin^2 (\Omega)}{h^5}-\dfrac{27\cos ^2(\Omega)\sin(\Omega)}{32 h^3}\nonumber\\&+\dfrac{9 \cos(\Omega)(31-\cos(2\Omega))\cot(\Omega)}{4096 h}\bigg)
+\dfrac{\sin ^3(\Omega)}{4 h_0 ^4}-\dfrac{27 \cos ^2(\Omega)\sin(\Omega)}{64 h_0^2}\nonumber\\&-\dfrac{9 \cos(\Omega)\cot(\Omega)(31-\cos(2\Omega))\log h_0}{4096}.
\end{align}
For  crease $ k\times R^2 $ we have
\begin{align}
S_{k\times R^2}=S_{(0,k\times R^2)}+S_{(1,k\times R^2)}
\end{align}
where
\begin{align}
&S_{(0,k\times R^2)}=-\frac{4\pi L^4\tilde{H}^2}{l_p ^4}\Bigg[-\frac{1}{2\delta ^2} \int_ {h_0}^{0}( \frac{dh}{\dot{h}}\dfrac{\sqrt{1+h^2 +\dot{h}^2}}{ h^2}+\frac{1}{h^2})-\frac{H}{3\delta ^3}+\frac{1}{2h_0 \delta ^2}\Bigg],
\label{c8}
\end{align}
and 
\begin{align}
&S_{(1,k\times R^2),0<\alpha<3/2, \alpha \ne 1}= \nonumber\\
&-\frac{4\pi L^4\tilde{H}^2 \mu ^{2\alpha}}{l_p ^4}\bigg[\bigg(\dfrac{ (3 + \alpha)}{2 h_0 (-1 + \alpha) (5 + 2 \alpha)}-\dfrac{4 h_0  }{(-1 + \alpha) (35 + 59 \alpha + 28 \alpha^2 + 
    4 \alpha^3)}\quad \nonumber\\
&-\dfrac{h_0^{-1 - 2 \alpha}
 a_1 \left(126 - 18 h_0^2 (-4 - 9 \alpha + 2 \alpha^2)+h_0^4 (-24 + 5 \alpha+ 107 \alpha^2 - 44 \alpha^3 + 
 4 \alpha^4)\right)}{
 252 (-1 + \alpha)}\bigg) \nonumber\\
&\times\dfrac{1}{\delta^{2 -  2 \alpha}}+\dfrac{H  }{(6 - 4 \alpha)\delta^{3 - 2 \alpha}}\nonumber\\
&+\frac{1}{2 (-1 + \alpha) \delta^{2 - 2 \alpha}}\int _{h_0}^0 dh\bigg(\dfrac{-
\dot{h} ^2- h^2 - 2 h^2\dot{h}  \dot{g_2}
+ 2 (2 + 2 \alpha+ h^2) g_ 2 h}{2 h^2 \dot{h}\sqrt{1+h^2 +\dot{h}^2}}-\big(\dfrac{3+2 \alpha}{ h^2(5+2\alpha)}\nonumber\\
&+\dfrac{8}{(35 + 59 \alpha + 28 \alpha ^2 + 4 \alpha ^3)}+ \dfrac{a_1 h^{-2 (1 + \alpha)}}{7}\left(-7 (1 + 2 \alpha) +  h^2 (4 + \alpha - 20 \alpha^2 + 4 \alpha^3\right))\big)\bigg)\bigg],
\end{align}

\begin{align}
&S_{(1,k\times R^2),\alpha=1}=\nonumber\\&
-\frac{4\pi L^4\tilde{H}^2\mu^{2 \alpha}}{l_p ^4}\bigg[\frac{H}{2 \delta}-\dfrac{(63 a_1 + 4 h_0^4 (1 + 6 a_1) + 
   9 h_0^2 (-4 + 11 a_1)}{63 h_0^3}\log(\mu \delta)\nonumber\\
  & +\log(\mu \delta)\int _{h_0}^0 dh\bigg(\dfrac{-
\dot{h} ^2- h^2 - 2 h^2\dot{h}  \dot{g_2}
+ 2 (4 + h^2) g_ 2 h}{2 h^2 \dot{h}\sqrt{1+h^2 +\dot{h}^2}}-\big(\dfrac{4}{63}+\dfrac{4-11 a_1}{7h^2}-\frac{3 a_1}{h^4}\big)\bigg)\bigg],\\
&S_{(1,k\times R^2),\alpha=\frac{3}{2}}=\frac{4\pi L^4\tilde{H}^2\mu ^{2\alpha}}{l_p ^4}\frac{
	H}{2}  \log(\mu \delta),\\
&S_{(1,k\times R^2), \alpha >3/2}=0.
\end{align}
\\
For entanglement entropy of $(c_1\times R^2)$ we have
\begin{align}
S_{(c_1\times R^2)}=S_{(0,c_1\times R^2)}+S_{(1,c_1\times R^2)},
\end{align}
where
\begin{align}
&S_{(0,c_1\times R^2)}=\nonumber\\&-\frac{4\pi ^2 L^5 \tilde{H}^2}{l_p ^5}\bigg[-\dfrac{\sin(\Omega) H^2}{8 \delta ^4}+\dfrac{\sin(\Omega)}{4 \delta ^2 h_0^3}-\dfrac{3 \cos(\Omega)\cot(\Omega)}{64}\frac{\log(\delta /H)}{\delta ^2}-\dfrac{3 \cos(\Omega)\cot(\Omega)}{64 \delta ^2}\nonumber\\
&\times(\frac{1}{2}+\log(h_0))-\frac{1}{2\delta ^2}\int _{h_0}^0 dh\big(\dfrac{\sin(\theta)\sqrt{1+h^2 +\dot{h}^2}}{\dot{h}h^3}+\dfrac{\sin(\Omega)}{h^3}-\dfrac{3 \cos(\Omega)\cot(\Omega)}{32 h}\nonumber\\
&-\dfrac{3(-13 +19 \cos(2\Omega))\cot ^2(\Omega)\sec(\Omega)h}{4096}\big)+\dfrac{3(-13 +19 \cos(2\Omega))\cot ^2(\Omega)\sec(\Omega)}{8192 H^2}\log(\delta/H)\bigg],
\end{align}
\begin{align}
&S_{(1,c_1\times R^2),0<\alpha<2,\alpha\neq1}=\nonumber\\&-\frac{4\pi ^2 L^5 \tilde{H}^2\mu^{2\alpha}}{l_p ^5}\bigg[-\dfrac{\sin(\Omega) H^2}{(16 - 8 \alpha) \delta ^{4-2\alpha}}+\dfrac{h_0^{-2 (1 + \alpha)}}{512 (-2 + \alpha) (-1 + \alpha)^2 (1 + \alpha) (2 + \alpha) \delta ^{2-2\alpha}}\nonumber\\
&\times\Bigg(768 h_0^{2 \alpha}- 1024 h_0^4 a_1 - 384 h_0^{2 \alpha} \alpha + 
   15 h_0^{2 + 2 \alpha} \alpha - 512 h_0^{(2 \alpha} \alpha^2 + 
   5 h_0^{2 + 2 \alpha} \alpha^2 + 1280\nonumber\\ &\times h_0^4 a_1 \alpha^2 + 64 h_0^{2 \alpha }\alpha ^3 + 64 h_0^{2 \alpha}\alpha^4 - 
   256 h_0^4 a_1 \alpha^4 - 
   5 h_0^{2 + 2 \alpha} (-13 + \alpha) \alpha \csc^2(\Omega) + 
   h_0^{2 + 2 \alpha}
     \nonumber\\ &\times\cot^2(\Omega)
      \bigg(5 (-24 + \alpha+ 7 \alpha^2) + 
      4 (36 - 28 \alpha - 19 \alpha^2 + 8 \alpha^3 +  3 \alpha ^4) \log(h_0) -  4 (36 \nonumber\\ &- 28 \alpha - 19 \alpha^2  + 8 \alpha^3 + 
         3 \alpha^4) \log(\delta/H) \bigg)\Bigg)\sin(\Omega)+\dfrac{1}{2 (-1 + \alpha)\delta ^{2-2\alpha}}\int_{h_0}^0 dh\nonumber\\
         & \Bigg(\dfrac{\sin(\theta) \left(-
\dot{h}^2 - h^2 (1 + 2 \dot{h}\dot{g_2}) +4 h^3 g_2 +  2 h (3 + 2 \alpha+ \dot{h}^2)g_2\right)}{2h^3\dot{h}\sqrt{1+h^2 +\dot{h}^2}}\nonumber\\
&-\bigg(\dfrac{(-18 + 5 \alpha + 3 \alpha^2) 
  \cos(\Omega) \cot(\Omega)}{64 h (-2 - \alpha + \alpha^2)}\nonumber\\&-\dfrac{h (-468 - 656 \alpha+ 332 \alpha^2 + 
   231 \alpha^3 - 
   39 \alpha^4)\cot^2(\Omega) \csc(\Omega) }{8192 (4 - 5 \alpha^2 + \alpha^4)}\nonumber\\
  &-\dfrac {h(684 - 112 \alpha- 436 \alpha^2 + 
      7 \alpha^3 + 57 \alpha^4) \cos(2\Omega) \cot^2(\Omega) \csc(\Omega)}{8192 (4 - 5 \alpha^2 + \alpha^4)}\nonumber\\
      &-2 h^{1 - 2 \alpha}  a_1 (-1 + \alpha)
   \sin(\Omega)+\dfrac{ (-6 + \alpha^2) \sin(\Omega)}{2 h^3 (-2 - \alpha+ \alpha^2)} \bigg)\Bigg)\bigg],
\end{align}
\begin{align}
&S_{(1,c_1\times R^2),\alpha=1}=\nonumber\\&-\frac{4\pi ^2 L^5 \tilde{H}^2\mu^{2\alpha}}{l_p ^5}\bigg[\dfrac{H^2 \sin(\Omega)}{8 \delta ^2}+ \frac{1}{256}
 \big(\cos(\Omega) \cot(\Omega) \bigg((-9 + 20\log(h_0) \log(\mu\delta)\nonumber\\ & - 10 \log^2(\delta/H) \big)+ 32(\frac{5}{h_0^2} - 8 a_1) \log(\mu\delta) \bigg)\sin(\Omega)+\log (\mu\delta)\int_{h_0}^0 dh\nonumber\\
         & \bigg(\dfrac{\sin(\theta) \left(-
\dot{h}^2 - h^2 (1 + 2 \dot{h}\dot{g_2}) +4 h^3 g_2 +  2 h (5+ \dot{h}^2)g_2\right)}{2h^3\dot{h}\sqrt{1+h^2 +\dot{h}^2}}\nonumber\\
&-\dfrac{5  \big(4096 - 71 h^4 \cot^4(\Omega) + 
   h^2 \cot^2(\Omega) \left(-256 + 71 h^2 + 89 h^2 \csc^2(\Omega)\right)\big) \sin(\Omega)}{16384 h^3}\bigg)\bigg],\\
&S_{(1,c_1\times R^2),\alpha=2}=\frac{\pi ^2 L^5 \tilde{H}^2\mu^{2\alpha}}{l_p ^5} H^2  \log(\mu\delta) \sin(\Omega),\\
&S_{(1,c_1\times R^2),\alpha>2}=0,
\end{align}
\section{Some Results for Complexity}
\label{Entropy2}
We have performed similar calculations for higher dimensional cones to what we did in $d=4$. In the case of $c_2$ we find two family of divergent terms proportional to $\log \delta$ and $\log ^2\delta$ as
\begin{align}
&\mathcal{C}_{5,c_2}^{\log}=\dfrac{L^4 }{8 G_N}\log \left(\frac{\delta}{H}\right)
\Bigg(\int_{h_0} ^0 dh\left(\dfrac{\sin ^2(\theta)}{h^4 \dot{h}}-\dfrac{4 \cos ^2 (\Omega) \cot (\Omega)}{9 h}+\dfrac{2 \cos (\Omega) \sin(\Omega)}{3 h^3}\right) \nonumber\\&-\dfrac{\cos(\Omega)\sin(\Omega)}{3 h_0 ^2}\Bigg)+\mathcal{C}_{5,c_2,\alpha}^{\log},\\
&\mathcal{C}_{5,c_2,\alpha=1}^{\log}=\dfrac{L^4\mu ^{2\alpha} }{8 G_N}\left[-\dfrac{H^2  (-9 + \alpha + 
   2 \alpha^2) \cos(\Omega) \sin(\Omega)}{6 (-3 - \alpha + 2 \alpha^2)}\right]\log\left(\mu\delta\right),\\
&\mathcal{C}_{5,c_2,\alpha=2}^{\log}=\dfrac{L^4 \mu ^{2\alpha}}{8 G_N}\left[\frac{1}{8} H^4  (2 \Omega - \sin(2 \Omega))\right]\log\left(\mu\delta\right),\\
&\mathcal{C}_{5,c_2}^{\log ^2}=\dfrac{L^4\mu ^{2\alpha}}{36 G_N}\cos ^2 (\Omega) \cot (\Omega) \log ^2\left(\dfrac{\delta}{H}\right).
\end{align}
For other values of $\alpha $ we don't have any logarithmic term. 
For the case of $c_3$ we also find 
\begin{align}
&\mathcal{C}_{6,c_3,\alpha=1/2}^{\log }=\dfrac{L^5\pi\mu ^{2\alpha}}{20 G_N}\log\left(\mu\delta\right)\Bigg[\dfrac{H  (3115 \cos(\Omega) + 473 \cos(3 \Omega)}{3072}\Bigg],\\
&\mathcal{C}_{6,c_3,\alpha=3/2}^{\log }=-\dfrac{L^5\pi\mu ^{2\alpha}}{20 G_N}\log\left(\mu \delta\right)\Bigg[\frac{3}{8} H^3  \cos(\Omega)\sin ^2(\Omega)\Bigg],\\
&\mathcal{C}_{6,c_3,\alpha=5/2}^{\log }=-\dfrac{L^5\pi\mu ^{2\alpha}}{20 G_N}\log\left(\mu \delta\right)\Bigg[-\frac{2}{3} H^5  \left(2 + \cos(\Omega)\right)\sin^4(\frac{\Omega}{2})\Bigg].
\end{align}
We have also performed the calculations for complexity of creases in higher dimensions. The result is as follows
\begin{align}
&\mathcal{C}_{(0,k\times R^d)}=\nonumber\\
&\dfrac{L^{d-1 }\tilde{H}^{d-3}}{8\pi  G}\Bigg[\dfrac{H^2 \Omega}{(d-1)\delta ^{d-1}}+ K_d  \left(-\dfrac{h_0}{d-3} (\dfrac{h_0}{\delta})^{d-3}+\dfrac{\delta}{d-2} (\dfrac{h_0}{\delta})^{(d-2)}\right) \nonumber\\
&-\dfrac{  1 }{(d-3)\delta ^{d-3}} \int _{h_0}^0  dq q^{d-3} J(q) \Bigg].
\end{align}
where
\begin{align}
&J(h)=\dfrac{1}{\dot{h}h^{d-1}}+ K _d,\\
&K_d=\dfrac{(1+h^2)^{\frac{(d-1)}{2}}}{h^{(d-1)} \sqrt{\dot{h}^2+h^2+1}}=\dfrac{(1+ h_0^2)^{\frac{(d-2)}{2}}} {h_0^{(d-1)}}.
\end{align}
\begin{align}
&\mathcal{C}_{(1,k\times R),0<\alpha<\frac{3}{2},\alpha\neq\frac{1}{2}}=\nonumber\\
&\dfrac{\tilde{H}L^3 \mu^{2\alpha}}{8\pi G}\Bigg[- h_0^{2 - 2 \alpha} K_4 \dfrac{
     12 a_1 (2 + \alpha) + 3 h_0^{2 \alpha} (-3 + 2 \alpha + \alpha^2) - 2 h_0^2 a_1(6 - 7 \alpha - \alpha^2 +    2 \alpha^3)}{ 12 (-1 + \alpha) (2 + \alpha) (-1 + 2 \alpha)\delta^{1 - 2 \alpha}}\nonumber\\
&+\dfrac{H^2 \Omega}{2 (-3 + 2 \alpha)\delta^{3 - 
  2 \alpha}} +\dfrac{ 1}{(1-2\alpha)\delta^{1-2\alpha}}\int _{h_0 }^{0 }dh\Bigg(\dfrac{1 - 4 h g_ 2 }{2  h^2 \dot{h}}\nonumber\\
  &-\bigg(\frac{1}{2} (-1 + 4 a_4) h K_4 - \frac{1}{4} (-3 + 12 a_4 - 8 a_5) h^3 K_4  +  h^{-2 \alpha}\big(h^3 K_4  (2 a_2 - 3 a_1) + 2 h K_4  a_1 \big)\bigg)\Bigg)\Bigg],
\end{align}
\begin{align}
&\mathcal{C}_{(1,k\times R),\alpha=1/2}=\nonumber\\
&\dfrac{\tilde{H}L^3 \mu^{2\alpha}}{8\pi G}\Bigg[-\dfrac{ H^2 \Omega}{4 \delta^2} - \frac{1}{60} h_0 K_4 \big(21 h_0 - 120 a_1 + 20 h_0^2 a_1\big) \log\big(\mu\delta\big)\nonumber\\
&- \log\big( \mu\delta\big)\int _{h_0 }^{0 }dh\bigg(\dfrac{1 - 4 h g_ 2 }{2  h^2 \dot{h}}-\big(\frac{1}{140} K_4 (-98 h + 171 h^3 + 280 a_1 - 140 h^2 a_1)\big)\bigg)\Bigg],\\
&\mathcal{C}_{(1,k\times R),\alpha=3/2}=\dfrac{ \tilde{H}H^2 L^3  \mu^{2\alpha}\Omega }{16 \pi G}\log(\mu \delta),\\
&\mathcal{C}_{(1,k\times R),\alpha>3/2}=0.
\end{align}
in which the constants $a_1-a_5$ are given in Eqs. (\eqref{h1}-\eqref{h5}). The complexity of crease $k \times R^2$ we find
\begin{align}
&\mathcal{C}_{(1,k\times R^2),0<\alpha<2,\alpha\neq1}=\nonumber\\
&\dfrac{ \mu^{2\alpha}L^4\tilde{H}^2}{8\pi G}\Bigg[\dfrac{ H^2  \Omega}{4(\alpha-2)\delta ^{4-2\alpha}}-\dfrac{1}{2(\alpha-1) \delta ^{2-2\alpha}}\bigg(\dfrac{h_{0} ^3 K_5 (5-\frac{60}{7}h_0 ^{2-2\alpha}a_1 (\alpha-2)+\frac{60 h_0 ^{-2\alpha }a_1}{2\alpha-3}+\frac{10}{5+2 \alpha})}{30}\nonumber\\
&+\int_{h_0}^0 dh\bigg[\dfrac{1-4 h g_2}{2 h^2 \dot{h}}-\frac{1}{14}K_5 h^2\bigg(-\dfrac{7(7+2\alpha)}{5+2 \alpha}-4 h^{-2\alpha}a_1\big(-7 + h^2 (10 - 9 \alpha+ 2 \alpha^2)\big)\bigg)\bigg]\Bigg)\Bigg],\\
&\mathcal{C}_{(1,k\times R^2),\alpha=1}=\nonumber\\ 
&\frac{\tilde{H}^2 L^4  \mu^{2\alpha}}{16 \pi G}\Bigg[-\dfrac{H^2 \Omega}{2 \delta ^2}+\log \left(\mu\delta\right)\bigg(-\frac{3}{7} h0^3 K_5 + 4 h_0 K_5 a_1 - \frac{4}{7}  h_0^3 K_5 a_1 +\nonumber\\
&\frac{1}{7}\int_{h_0}^0 dh\big(\dfrac{7(1-4 h g_2)}{h^2\dot{h}}+9 h^2 K_5  + 4 (-7 + 3 h^2) K_5  a_1\big)\bigg)\Bigg],\\
&\mathcal{C}_{(1,k\times R^2),\alpha=2}=\dfrac{H^2 L^4 \tilde{H}^2  \mu^{2\alpha} \Omega}{16 \pi G}\log \left(\mu\delta\right),\\
&\mathcal{C}_{(1,k\times R^2),\alpha>2}=0.
\end{align}
For crease $c_1 \times R$ the corresponding subregion complexity  turns out to be as follows
\begin{align}
\mathcal{C}_{c_1\times R}=\mathcal{C}_{(0,c_1\times R)}+\mathcal{C}_{(1,c_1\times R)},
\end{align}
\begin{align} 
\mathcal{C}_ {(0,c_1\times R)}=&\nonumber\\
&-\dfrac{L^4 \tilde{H} \Omega _1}{32 \pi G}\Bigg[-\dfrac{H^3}{3 \delta ^4}( \cos(\Omega)-1)+\dfrac{\cos(\Omega) H}{6 \delta ^2}
\\&
-\frac{1}{\delta}\left(\dfrac{\cos(\Omega)}{3 h_0}+\dfrac{(-13+5 \cos (2\Omega)) \cot (\Omega) \csc(\Omega) h_0}{108 }-\int _ {h_0}^{0} dq q J_5(q) \right)\\ &-\dfrac{(-13+5 \cos (2\Omega)) \cot (\Omega) \csc(\Omega) }{108 H}\log \left(\frac{\delta}{H}\right)\Bigg],
\end{align}
where
\begin{align}
J_5(h)=\dfrac{\sin(\theta)}{\dot{h} h^4}+\dfrac{\cos (\Omega)}{3 h^3}-\dfrac{(-13+5 \cos(2 \Omega) ) \cot (\Omega) \csc (\Omega )}{108 h}.
\end{align}
and
\begin{align}
&\mathcal{C}_{(1,c_1\times R),0<\alpha<2,\alpha \neq\frac{1}{2},1}=\nonumber\\
&\dfrac{\mu^{2\alpha}L^4 \tilde{H}}{12 G}\Bigg[-\dfrac{H^3 \left(-1 + \cos(\Omega)\right)}{4 (-2 + \alpha) \delta^{4-2\alpha}}-\dfrac{H  (-9 + \alpha + 2 \alpha^2) \cos(\Omega)}{12 (-1 + \alpha) (-3 - \alpha + 2 \alpha^2) \delta^{2-2\alpha}}\nonumber\\
&-\dfrac{h_0^{-(1 + 2 \alpha)}
   \left(3 h_0^3 a_1 -\dfrac{  2 h_0^{2 \alpha }(6 - 7 \alpha + \alpha^3)}{-3 - \alpha + 
     2 \alpha^2}\right) \cos(\Omega)}{6 (-1 + \alpha) (-1 + 2 \alpha) \delta^{1-2\alpha}}+\dfrac{1}{(1-2\alpha)\delta^{1-2\alpha}}\nonumber\\
     &\times\int_{h_0}^0 dh \Bigg(-\dfrac{\sin(\theta)(-1+6 h g_2)}{2\dot{h}h^3}-\bigg(h^{1 - 2 \alpha} a_1  \cos(\Omega) - \dfrac{(-6 + \alpha + \alpha^2) \cos(\Omega)}{3 h^2 (-3 - \alpha + 2 \alpha^2)}\bigg)\Bigg)\Bigg],
\end{align}

\begin{align}
&\mathcal{C}_{(1,c_1\times R),\alpha=\frac{1}{2}}=\nonumber\\
&\dfrac{ \mu^{2\alpha}L^4 \tilde{H}}{12G}\Bigg[\dfrac{4 H \cos(\Omega)}{9 \delta}+\dfrac{7 \cos(\Omega) \log(\mu\delta)}{12 h_0}  + 
 h_0 a_1 \cos(\Omega) \log(\mu\delta) - \dfrac{H^3 \sin^2(\frac{\Omega}{2})}{3 \delta^3}\nonumber\\
 &+\log(\mu\delta)\int_{h_0}^0 dh \Bigg(-\dfrac{\sin(\theta)(-1+6 h g_2)}{2\dot{h}h^3}-\bigg(h^{1 - 2 \alpha} a_1  \cos(\Omega) -\dfrac{7\cos(\Omega)}{12 h^2}\bigg)\Bigg)\Bigg],\\
&\mathcal{C}_{(1,c_1\times R),\alpha=1}=\dfrac{ \mu^{2\alpha}L^4 \tilde{H}}{12 G}\Bigg[-\frac{1}{2}H \cos(\Omega) \log(\mu\delta) - \dfrac{H^3\sin^2\left(\dfrac{ \Omega}{2}\right) }{2 \delta^2} \Bigg],\\
&\mathcal{C}_{(1,c_1\times R),\alpha=2}=\dfrac{ \mu^{2\alpha}L^4 \tilde{H}}{12 G}\left[H^3\sin^2\left(\dfrac{ \Omega}{2}\right)\log(\mu\delta)\right],\\
&\mathcal{C}_{(1,c_1\times R),\alpha>2}=0.
\end{align}

\end{document}